\documentclass{jfm}
\usepackage{graphicx}
\usepackage{amsmath}
\usepackage[dvipsnames]{xcolor}

\usepackage{multirow}
\usepackage{bm,upgreek}

\shorttitle{Statistics, plumes and waves in ultimate TC turbulent vortices}
\shortauthor{A. Froitzheim et al.}

\title{{Statistics, plumes and azimuthally traveling waves in ultimate Taylor-Couette turbulent vortices}}

\author{Andreas Froitzheim\aff{1},
  Rodrigo Ezeta\aff{2},
  Sander G. Huisman\aff{2},
  Sebastian Merbold\aff{1},
  Chao Sun\aff{3,2},
  Detlef Lohse\aff{2,4}
 \and {Christoph~Egbers}\aff{1}
 \corresp{\email{christoph.egbers@b-tu.de}}
 }

\affiliation{\aff{1}Department of Aerodynamics and Fluid Mechanics, Brandenburg University of Technology Cottbus-Senftenberg, Siemens-Halske-Ring 14, 03046 Cottbus, Germany
\aff{2}Physics of Fluids Group, {Max Planck Center Twente for Complex Fluid Dynamics}, MESA+ Institute and J.M. Burgers Centre of Fluid Dynamics, {Department of Science \& Technology}, University of Twente, P.O. Box 217, 7500AE Enschede, The Netherlands
\aff{3}Center for Combustion Energy and Department of Thermal Engineering, Tsinghua University, Beijing 100084, China
\aff{4}Max Planck Institute for Dynamics and Self-Organisation, {Am Fa\ss berg 17}, 37077 G\"{o}ttingen, Germany}

\begin{document}

\maketitle

\begin{abstract}
In this paper{,} we experimentally study the influence of large-scale Taylor rolls on the small-scale statistics and the flow organization in fully turbulent Taylor-Couette flow {for Reynolds numbers up to $\Rey_S=3\times 10^5$}. The velocity field in the gap confined by coaxial and independently rotating cylinders at a radius ratio of $\eta=0.714$ is measured using planar {particle image velocimetry} in horizontal planes at different cylinder heights. Flow regions with and without prominent Taylor vortices are compared. We show that the local angular momentum transport (expressed in terms of a Nusselt number) mainly takes place in the regions of the vortex in- and outflow, where the radial and azimuthal velocity components are highly correlated. The efficient momentum transfer is reflected in intermittent bursts, which becomes visible in the exponential tails of the probability density functions of the local Nusselt number. In addition, by calculating azimuthal energy co-spectra, small-scale plumes are revealed to be the underlying structure of these bursts. These flow features are very similar to the one observed in Rayleigh-B\'{e}nard convection, which emphasizes the analogies of these both systems. By performing a {complex proper orthogonal decomposition}, we remarkably detect azimuthally traveling waves superimposed on the turbulent Taylor vortices, not only in the classical but also in the ultimate regime. This very large-scale flow pattern{,} which is most pronounced at the axial location of the vortex center, is similar to the well-known wavy Taylor vortex flow{,} which has comparable wave speeds, but much larger azimuthal wave numbers.
\end{abstract}


\section{Introduction}
The flow in between two independently rotating cylinders, known as Taylor-Couette (TC) flow, is a commonly used model for general rotating shear flows. It features rich and diverse flow states, which have been explored for nearly a century \citep{Taylor1923,wen33,Coles1964,Andereck86,Donnelly1991}. More recent reviews on the hydrodynamic instabilities can be found in \citet{Fardin2014}, and on fully turbulent Taylor-Couette flows in \citet{Grossmann2016}.
The geometry of a Taylor-Couette system is defined by the gap width $d=r_2-r_1$, where $r_1$ and $r_2$ are the inner and outer radii respectively; the radius ratio $\eta=r_1/r_2$, and the aspect ratio $\Gamma=\ell/d$, with $\ell$ the height of the cylinders. The external driving of the flow can be quantified by the shear Reynolds number according to \citet{Dub2005}

\begin{equation}
\Rey_S=\frac{2r_1r_2d}{(r_1+r_2) \nu}\vert \omega_2-\omega_1 \vert =\frac{u_S d}{\nu},
\end{equation}

\noindent with the cylinder angular velocities $\omega_{1,2}$, $\nu$ is the kinematic viscosity of the fluid, and $u_S$ is the shear velocity. A further dimensionless control parameter is the ratio of the angular velocities

\begin{equation}
\mu=\frac{\omega_2}{\omega_1},
\end{equation}

\noindent implying $\mu>0$ for co-rotation of the cylinders, $\mu=0$ for pure inner cylinder rotation and $\mu<0$ for counter-rotation. The most important response parameter of the TC system to the cylinder driving is the angular velocity transport \citep{Eckhardt07b}

\begin{equation}
J_\omega=r^3 \left( \langle u_r \omega \rangle_{\varphi,z,t} -\nu \partial_r \langle \omega \rangle_{\varphi,z,t} \right),
\end{equation}

\noindent where $r$ denotes the radial coordinate, $\varphi$ the azimuthal coordinate, $t$ the time coordinate and $\langle \cdot   \rangle_{\varphi,z,t}$ the azimuthal-axial-time average. This quantity is conserved along $r$ and can be directly measured by the torque $\mathcal{T}$ acting on either the inner (IC) or the outer cylinder wall (OC). Normalizing $J_\omega$ with its corresponding laminar non-vortical value ${J_{\omega}^{lam}=2\nu r_1^2r_2^2(\omega_1-\omega_2)/(r_2^2-r_1^2)}$ yields a quasi Nusselt number $\text{Nu}_\omega=J_\omega/J_{\omega}^{lam}$ \citep{Eckhardt07b}, which is analogous to the Nusselt number $\text{Nu}$ in Rayleigh B\'{e}nard flow (RB) flow, i.e. the buoyancy{-}driven flow which is heated from below and cooled from above. There, $\text{Nu}$ is a measure for the amount of transported heat flux normalized by the purely conductive heat transfer. \citet{Eckhardt07a} worked out the fundamental similarities between TC and RB flow in terms of the Nusselt number and the energy dissipation rate, which we will use in this paper.

The dependence of the Nusselt number on the shear Reynolds number, commonly expressed in terms of an effective power law $\text{Nu}_\omega \sim \Rey_S^\alpha$, and on the rotation ratio $\mu$ has been widely investigated \citep{Lathrop92,Lewis99,vanGils2011b,Paoletti11,vanGils2012,Merbold13,Ostilla14b,Brauckmann2016,Grossmann2016}. For pure inner cylinder rotation ($\mu=0$), a change in the local scaling exponent $\alpha$ with increasing $\Rey_S$ has been found which is caused by a transition from laminar (classical regime) to turbulent boundary layers (ultimate regime) \citep{Ostilla14b}. The transition point depends on the radius ratio $\eta$ and is located around $Re_{S,crit}\approx 1.6\times 10^4$ for $\eta=0.714$. In the ultimate regime, the scaling exponent becomes $\alpha \approx 0.76$ independent on $\eta$ \citep{Ostilla13}. {When the driving ($\Rey_S$) is kept constant and only $\mu$ is changed, the TC flow features a maximum in the angular momentum transport ($\text{Nu}_\omega$)}. For $\eta=0.714$, the maximum is located in the counter-rotating regime at $\mu_{max}\approx -0.36$ and originates from a strengthening of the turbulent Taylor vortices \citep{Brauckmann13b,Ostilla14c}. {The contribution of these rolls to the angular momentum transport has been evaluated numerically by \citet{Brauckmann13b} and experimentally by \citet{Froitzheim2017} by means of the decomposition of  $\text{Nu}_\omega$ into its turbulent fluctuation and large-scale circulation contributions. They find that the large-scale contribution dominates the transport in the region of the torque maximum. {Besides, \citet{Tokgoz2011} could show by performing direct torque measurements and tomographic PIV measurements in a TC facility at $\eta=0.917$ and $Re_S \in [1.1\times 10^4;  2.9\times 10^4]$ that the torque is strongly affected by the rotation ratio, which determines whether large-scale or small-scale structures are dominant in the flow.} These findings reflect that turbulent Taylor vortices can play a prominent role in the fully turbulent regime.}

\begin{figure}
\begin{center}
\includegraphics[scale=0.7]{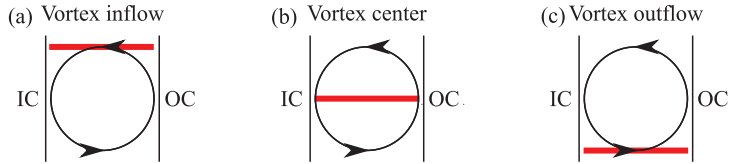}
\caption{Sketch of the ejecting regions bounded by the inner (IC) and outer (OC) cylinders within a Taylor roll: (a) vortex inflow, (b) vortex center, (c) vortex outflow.}
\label{fig:sketch_ejecting}
\end{center}
\end{figure}

Hence, the morphology and physical mechanisms behind these roll structures have been in the focus of different studies throughout the literature. An interesting phenomenon regarding Taylor vortices is the reappearance of azimuthal waves in the turbulent Taylor vortex regime. \citet{Walden1979} measured the point-wise radial velocity component close to the OC boundary layer for $\mu=0$ at $\eta=0.875$ and for different aspect ratios. They find a regime of reappearance for $28 \leq \Rey/\Rey_C \leq 36$ and for $\Gamma \geq 25$, based on sharp peaks in the power spectrum. Here, $\Rey_C$ is the critical Reynolds number for the onset of Taylor vortex flow. Later, \citet{Takeda1999} acquired time-resolved axial profiles of the axial velocity component using an ultrasonic measurement technique for $\eta=0.904$ and $\Gamma=20$. The azimuthal waves are identified based on Fourier analysis and {proper orthogonal decomposition} (POD) in the range of $23 \leq \Rey/\Rey_C \leq 36$. Their results show good agreement with those of \citet{Walden1979}. In another study, \citet{Wang2005} performed planar PIV measurements in a meridional plane for $\eta=0.733$ and $\Gamma=34$. They capture the reappearance of azimuthal waves for $20 \leq \Rey/\Rey_C \leq 38$ based on spatial correlations. Note that the three aforementioned studies are all based on pure inner cylinder rotation ($\mu=0$). More recently, \citet{Merbold2014} performed flow visualizations in TC flow with $\eta=0.5$ at $\Rey_S=5000$. They find an axial oscillation of the turbulent Taylor vortices in the range of $\mu \in [-0.15,-0.3]$, which includes the rotation ratio for optimum transport $\mu_{max}=-0.2$. In {summary}, the large-scale turbulent Taylor rolls seem to feature an instability mechanism similar to the one in the laminar regime, {which, however, has not yet been detected in highly turbulent TC flows}.
Based on numerical simulations, \citet{Ostilla13,Ostilla14b} further showed that the large-scale rolls consist and are driven by small-scale unmixed plumes, in analogy to RB flow. They calculated {for $\mu=0$ and $\eta=0.714$} the radial profiles of the angular velocity $\omega$ at specific axial positions of the large-scale Taylor vortices; namely at the vortex inflow, vortex center and vortex outflow. The vortex inflow is characterized by the ejection of plumes from the OC in conjunction with a mean radial velocity that points away from the OC (see the sketch in figure \ref{fig:sketch_ejecting}a). In contrast, the outflow features plume ejections from the IC with a mean radial velocity component directed from the IC to the OC (see the sketch in figure \ref{fig:sketch_ejecting}c). In between the in- and outflow, the radial velocity component becomes zero in the middle of the gap, which denotes the location of the center of the vortex, as shown in figure \ref{fig:sketch_ejecting}b. In regions where plumes are ejected from the cylinder walls, i.e. in the vortex inflow at the outer wall and in the vortex outflow at the inner wall, the radial profiles of $\omega$ have a logarithmic shape in the corresponding boundary layer \citep{Ostilla2016}. {It is worth to mention that in the ultimate regime, the profiles become logarithmic also in the absence of dominant large-scale rolls \citep{hui13}}. For $\eta=0.5$ {in the classical regime}, \citet{vanderVeen2016a} calculated the velocity of such plumes based on planar particle image velocimetry (PIV) measurements performed at different heights, which  further confirms the connection between small-scale plumes and large-scale vortices.

Within the context of both TC and RB flow, many studies in the literature focus on establishing a distinct connection between the transport of angular momentum (or heat in RB flow) and the structures inherent to these flows, both for large-scale rolls and small-scale plumes.  A comparative study of the probability density functions (PDFs) of the Nusselt number in TC and RB calculated over cylindrical surfaces and horizontal planes respectively has been performed by \citet{Brauckmann2016b}. As a reference point for the comparison, they choose $\Rey_S=2\times 10^4$ for TC and a Rayleigh number of $Ra=\alpha_P gH^3 \Delta T / (\kappa \nu)=10^7$ for RB, where the Nusselt number is identical for both flows. Here, $Ra$ is the dimensionless driving parameter in RB flow with $\alpha_p$ the thermal expansion coefficient, $g$ the gravitational acceleration, $H$ the height of the RB cell, $\Delta T$  the temperature difference, and $\kappa$ the thermal diffusivity. They find that the PDFs of the net transport have the same asymmetric shape with differences in the width of the tails in the boundary layer regions. These differences can be attributed to different shapes and detachment frequencies of the plumes. Moreover, the PDFs of the angular momentum and temperature fluctuations depict a cusp-like shape with pronounced exponential tails as a consequence of the effect of intermittent bursting plumes. 
Similar analyses of heat flux PDFs in RB flow have been performed by \citet{Shishkina2007}. They find that the instantaneous heat flux fluctuates around zero and not around the volume-averaged Nusselt number, along with a broadening of the tails with increasing $Ra$. \citet{Shang2004} revealed that the asymmetry of the PDFs of $\text{Nu}$, which mainly occurs in the tails, arises from correlated temperature and velocity signals produced by thermal plumes. These plumes lead to large but rare positive events of heat flux. However, the results of \citet{Shang2004} are only based on point-wise measurements. For TC flow, the statistics of $\text{Nu}_\omega$ were analyzed {for $\mu=0$ by \citet{Huisman2012}} without any connection to specific flow structures.

Another approach {investigating} structures in TC flow is to analyze the kinetic energy spectra. \citet{Dong07} performed direct numerical simulations (DNSs) for $\eta=0.5$, $\Rey=8000$ and the OC at rest. Strikingly, he finds a small-scale peak in the axial spectra of the radial velocity component. According to \citet{Dong07} the underlying structures can be specified as herringbone streaks. The DNSs of \citet{Ostilla2016} in the boundary layer regions reveal a peak in the azimuthal and axial spectra of the radial velocity component at large wavenumbers, which indicates the existence of small-scale plumes. Their simulations were done for $\eta=0.909$ and $\mu=0$ at $\Rey_S \geq 10^5$. We stress that the energy spectra in TC flow do not follow the classical Kolmogorov scaling for homogeneous and isotropic turbulence (HIT) of -5/3 \citep{Lewis99,Dong07,Hout2011,Huisman2013,Ostilla2016}. 

{Based on this literature review, the following open questions are addressed within this manuscript: how do turbulence-dominated and vortex-dominated flow states differ with respect to their velocity field statistics, how important are the vortex inflow, vortex center and vortex outflow regions for the momentum transport, how do small-scale structures affect this transport, what is the lengthscale of these structures and do Wavy-vortex-like turbulent Taylor vortices exist in the ultimate turbulent regime{?} To answer these questions, we make use of PIV measurements in horizontal planes at different cylinder heights for $\eta=0.714$, in the range of $\Rey_S \in [9.3 \times 10^3,3.5 \times 10^5]$ and $\mu \in [0,-0.36]$. We want to stress that such quasi-threedimensional experimental investigations of the local angular momentum transport statistics and flow structures in the ultimate turbulent TC flow at $\mu_{max}$ are unique, while the measurement setup consisting of a TC apparatus with a transparent top plate and a horizontal PIV configuration has already been used successfully by \citet{vanderveen2016b} and \citet{Froitzheim2017}.}

{The paper is organized as follows. In $\S\,2$, the experimental setup, the measurement technique, and the investigated parameter space are described in detail. Thereafter, the flow states, statistical profiles, and velocity PDFs are shown and compared to the literature to prove the quality of the measurements and discuss the influence of large-scale turbulent Taylor vortices onto the global flow statistics ($\S\,3$). In $\S\,4$ the global and local angular momentum transport are analyzed based on the net convective Nusselt number and the contributions of the vortex in- and outflow to the overall transport are worked out. To detect intermittent bursting small-scale structures which influence this transport, the PDFs of the net convective Nusselt number are evaluated over cylindrical surfaces as well as at the axial height of the vortex inflow, center, and outflow in $\S\,5$. The energy content and azimuthal lengthscale of these structures are calculated in $\S\,6$ based on azimuthal energy co-spectra, while azimuthally traveling waves superimposed to the turbulent Taylor vortices are extracted from the flow field based on a complex proper orthogonal decomposition in $\S\,7$. The paper ends with a summary and a conclusion ($\S\,8$).}

\section{Experimental setup}

\begin{figure}
\begin{center}
\includegraphics[scale=0.5]{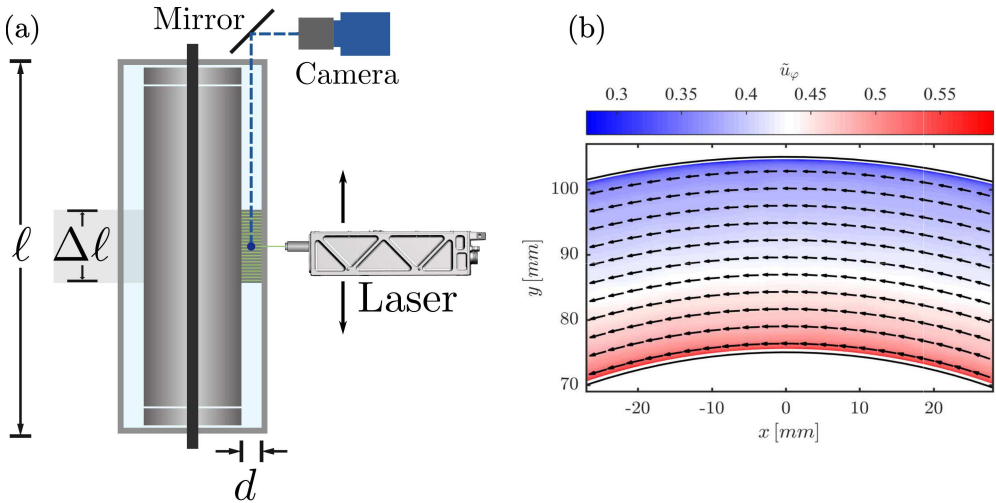}
\caption{(a) Sketch of the experimental apparatus (BTTC). Shown is a vertical section. A mirror set at 45 degrees is used so the camera captures the velocity field in the $r-\varphi$ plane. The picture shows an exaggeration of the 23 positions of the laser sheet along $\Delta  \ell$. (b) Temporally averaged flow field in a horizontal plane at the axial height of the vortex center for $\Rey_S=3.51 \times 10^5$ and $\mu=0$. Colors represent the azimuthal velocity component and arrows the velocity field. Only each tenth vector is shown. Black solid lines indicate the position of the inner and outer cylinder. 
}
\label{exp_setup}
\end{center}
\end{figure}

The PIV experiments were performed in the {b}oiling Twente Taylor-Couette facility (BTTC) at the University of Twente. The BTTC is an ideal facility to perform PIV experiments due to its transparent outer cylinder and top plate. The inner and outer radius of the setup are $r_1=75 \, \textrm{mm}$ and $r_2=105\, \textrm{mm}$, respectively and thus the radial gap is $d=r_2-r_1=30 \, \textrm{mm}$. The height of the cylinders is $\ell=549 \, \textrm{mm}$, which gives an aspect ratio of $\Gamma=\ell/d=18.3$. The radius ratio is then $\eta=r_1/r_2=0.714$. A more detailed overview of the setup can be found in \cite{Huisman2015}. The flow consists of water whose viscosity $\nu$ and density $\rho$ can be controlled throughout the experiments due to the temperature control of the BTTC. We fix the temperature of the experiments to $20 \, ^{\circ}\text{C}$, leading to $\nu=1.002 \, \textrm{mm}^2\textrm{/s}$ and $\rho=0.998 \, \textrm{g/cm}^3$. The standard deviation of the temperature is $15 \, \textrm{mK}$. The flow is seeded with fluorescent polyamide particles with diameters up to $20 \, \mu\textrm{m}$ with an average particle density of approximately $0.01 \, \textrm{particles}/\textrm{pixel}$. These particles are coated with Rhodamine B which has a maximum emission centered at around $565 \, \textrm{nm}$.  The illumination is provided by a laser sheet from a {double-pulsed} cavity laser (Quantel Evergreen 145 laser, $532 \, \textrm{nm}$). The thickness of the laser sheet is $\approx 1 \, \textrm{mm}$. The laser is mounted to a traverse system (Dantec lightweight traverse) which allows us to precisely change the location of the laser sheet along the vertical direction. The camera we used for the recordings is an Imager sCMOS ($2560\times2160 \, \textrm{px}$) 16 bit with a Carl Zeiss Milvus 2.0/100 lens. We capture the velocity fields with a framerate of $f=15 \, \textrm{Hz}$. Since the camera is operated in double frame mode, we can have very small interframe times, i.e. $\Delta t\ll 1/f $. In order to maximize the contrast of the images, we use a long-pass filter in front of the lens (Edmund {High-Performance} Longpass Filter, $550 \, \text{nm}$), which collects only the emitted light from the fluorescent particles. In figure \ref{exp_setup}a, we show a sketch of the experimental setup.

The velocity fields are calculated with commercial software (Davis 8.0) using a multi-pass method. The algorithm uses windows of size $64\times 64 \, \textrm{px}$ for the first pass and windows of $24\times 24 \, \textrm{px}$ for the last iteration with a 50\% overlap of the windows. This process yields the velocity fields in Cartesian coordinates. In order to have access to the velocity fields in polar coordinates, we map the Cartesian velocity fields onto a polar grid using bilinear interpolation. The mapping is done such that the radial $\Delta r$ and azimuthal $\Delta \varphi$ resolution is the same as the spatial resolution in Cartesian coordinates $\Delta x$, i.e. $\Delta r=\Delta x$ and $r \Delta \varphi=\Delta x$. In this way, the resultant velocity fields are of the form ${\bf u}=u_r(r,\varphi,t){\bf e_r}+u_\varphi(r,\varphi,t){\bf e_\varphi}$, where $u_r$ and $u_\varphi$ are the radial and azimuthal velocity components which depend on the radial coordinate $r$, the azimuthal coordinate $\varphi$ and the time coordinate $t$. ${\bf e_r}$ and ${\bf e_\varphi}$ are the unit vectors in the radial and azimuthal direction respectively.

In table \ref{table:points}, we present a summary of the measurements we performed. In total, 8 cases were investigated which will be addressed in the following sections. {Each case contains measurements done at 23 different heights which are separated by $\Delta z =4\, \textrm{mm}$, and each height contains 1500 different velocity fields.} Thus, the height of the experiments spans a length of $\Delta \ell=22\Delta z=88 \, \textrm{mm}$. The resolution of the velocity fields  $\Delta x$ depends on the height but lies within $\Delta x\in[0.607,0.752]\ \textrm{mm}$, where the smallest value corresponds to the height closest to the camera at $(z-\ell/2)/d=1.5$ and the smallest to $(z-\ell/2)/d=-1.5$.

We investigate flow states for pure inner cylinder rotation as a reference for a turbulence-dominated flow and for the rotation ratio that corresponds to the torque maximum, where pronounced large-scale Taylor rolls are present in the gap (see table \ref{table:points}). {Further, the measurements are classified based on the discovered change in the local scaling exponent $\alpha$ by \cite{Ostilla14c} at $\Rey_S(\eta=0.714) \approx 1.6 \times 10^4$ for $\mu=0$, which is caused by a transition of the boundary layers (BLs). Accordingly, flow states at $\Rey_S < 1.6 \times 10^4$ are assumed to be in the so-called classical regime with laminar BLs, while those at $\Rey_S > 1.6 \times 10^4$ are assumed to be in the ultimate regime with turbulent BLs. Case 1 and 2 are in the classical regime, where, $\mu_{max}$ changes with the shear Reynolds number, which is why $\mu_{max}=-0.15$ is different to the cases in the ultimate regime \citep{Ostilla13}. For the flow states in the ultimate regime, the torque maximum is located around $\mu_{max}=-0.36$.}

\begin{table}
\renewcommand{\arraystretch}{1.5}
  \begin{center}
\def~{\hphantom{0}}
  \begin{tabular}{cccccc}
 Case & Regime & $\Rey_S$ & $\mu$ & Abbreviation $\text{C}_{\#}\vert_{Re_S}^{\mu}$ & Linestyle \\
1 & classical & $9.32\times 10^3$ & $0$ & $\text{C}_1\vert _{9.32\times10^3}^{0}$ & \textbf{\textcolor[rgb]{0,0.447,0.741}{-\hspace{0.04cm}-\hspace{0.04cm}-}}\\
2 & classical & $9.30\times 10^3$ & $-0.15$ & $\text{C}_2\vert _{9.30\times10^3}^{-0.15}$ & \textbf{\textcolor[rgb]{0,0.447,0.741}{-\hspace{-0.05cm}-\hspace{-0.04cm}-\hspace{-0.04cm}-\hspace{-0.04cm}-}}\\
3 & ultimate & $2.98\times 10^4$ & $-0.36$ & $\text{C}_3\vert _{2.98\times10^4}^{-0.36}$ & \textbf{\textcolor[rgb]{0.301,0.745,0.9330}{-\hspace{-0.05cm}-\hspace{-0.04cm}-\hspace{-0.04cm}-\hspace{-0.04cm}-}}\\
4 & ultimate & $6.68\times 10^4$ & $-0.36$ & $\text{C}_4\vert _{6.68\times10^4}^{-0.36}$ & \textbf{\textcolor[rgb]{0.466,0.674,0.1880}{-\hspace{-0.05cm}-\hspace{-0.04cm}-\hspace{-0.04cm}-\hspace{-0.04cm}-}} \\
5 & ultimate & $9.46\times 10^4$ & $-0.36$ & $\text{C}_5\vert _{9.46\times10^4}^{-0.36}$ & \textbf{\textcolor[rgb]{0.929,0.694,0.125}{-\hspace{-0.05cm}-\hspace{-0.04cm}-\hspace{-0.04cm}-\hspace{-0.04cm}-}} \\
6 & ultimate & $2.15\times 10^5$ & $0$ & $\text{C}_6\vert _{2.15\times10^5}^{0}$ & \textbf{\textcolor[rgb]{0.85,0.325,0.098}{-\hspace{0.04cm}-\hspace{0.04cm}-}}\\
7 & ultimate & $2.14\times 10^5$ & $-0.36$ & $\text{C}_7\vert _{2.14\times10^5}^{-0.36}$ & \textbf{\textcolor[rgb]{0.85,0.325,0.098}{-\hspace{-0.05cm}-\hspace{-0.04cm}-\hspace{-0.04cm}-\hspace{-0.04cm}-}} \\
8 & ultimate & $3.51\times 10^5$ & $0$ & $\text{C}_8\vert _{3.51\times10^5}^{0}$ & \textbf{\textcolor[rgb]{0.635,0.078,0.184}{-\hspace{0.04cm}-\hspace{0.04cm}-}}
  \end{tabular}
  \caption{Overview of investigated flow states, defined by the regime, the shear Reynolds number $Re_S$, and the rotation ratio $\mu$. The penultimate column depicts the abbreviations for the different flow states and the last column depicts the linestyles used in the study. The transition point from the classical to the ultimate regime is located at $\Rey_S \approx 1.6 \times 10^4$.}
  \label{table:points}
  \end{center}
\end{table}

\section{Flow states and velocity profiles}

\begin{figure}
  \centerline{\includegraphics[width=\textwidth]{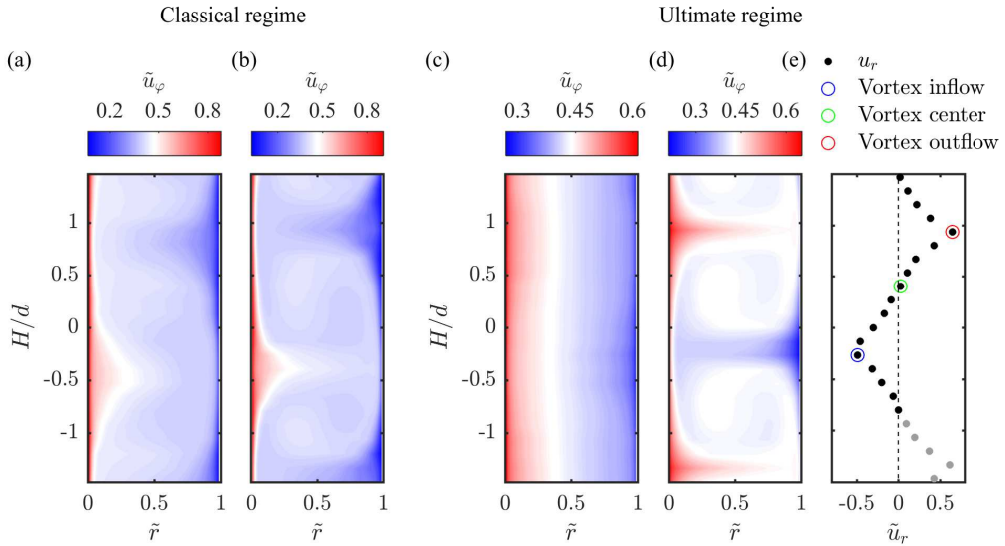}}
  \caption{Representation of the flow states in terms of temporal and azimuthally averaged velocities. The contour plots depict the azimuthal velocity component  for {(a) $Re_S=9.30\times10^3$ and $\mu=0$ ($\text{C}_1$), (b) $9.30\times 10^3$ and $\mu=-0.15$ ($\text{C}_2$), (c) $Re_S=2.15\times 10^5$ and $\mu=0$ ($\text{C}_6$) and (d) $Re_S=2.14\times 10^5$ and $\mu=-0.36$ ($\text{C}_7$). (e) Axial profile of the radial velocity component for $Re_S=2.14\times 10^5$ and $\mu=-0.36$ ($\text{C}_7$)} {at $\tilde{r}=0.5$} with marked location of vortex inflow, vortex center and vortex outflow. Gray dots represent data points, which are excluded for flow quantities calculated over the axial coordinate.}
\label{fig:flow_states}
\end{figure}

{The TC flow at $\mu_{max}$ is dominated by turbulent Taylor vortices, while at $\mu=0$ a featureless turbulent  flow state develops inside the gap \citep{vanGils2011b,Brauckmann13b,Ostilla13,Huisman2014}. In the following, this previous finding is confirmed within our measurements and statistical profiles and velocity PDFs over cylindrical surfaces are compared to the literature for validation. Therefore, the azimuthally and time averaged azimuthal velocity components for the 23 investigated heights are shown in figure \ref{fig:flow_states}. The tilde symbols denote normalized quantities. The radial coordinate is normalized as $\tilde{r}=(r-r_1)/(r_2-r_1)$, where 0 means the location of the inner cylinder ($r_1$) and 1 the location of the outer cylinder ($r_2$). Further, the azimuthal velocity is normalized using the cylinder speeds as $\tilde{u}_\varphi=(u_\varphi-u_{\varphi,2})/(u_{\varphi,1}-u_{\varphi,2})$, with $u_\varphi=r \omega$. In order to normalize the radial velocity component, the shear velocity is used $\tilde{u}_r=u_r/u_S$.}

\begin{figure}
  \centerline{\includegraphics[width=\textwidth]{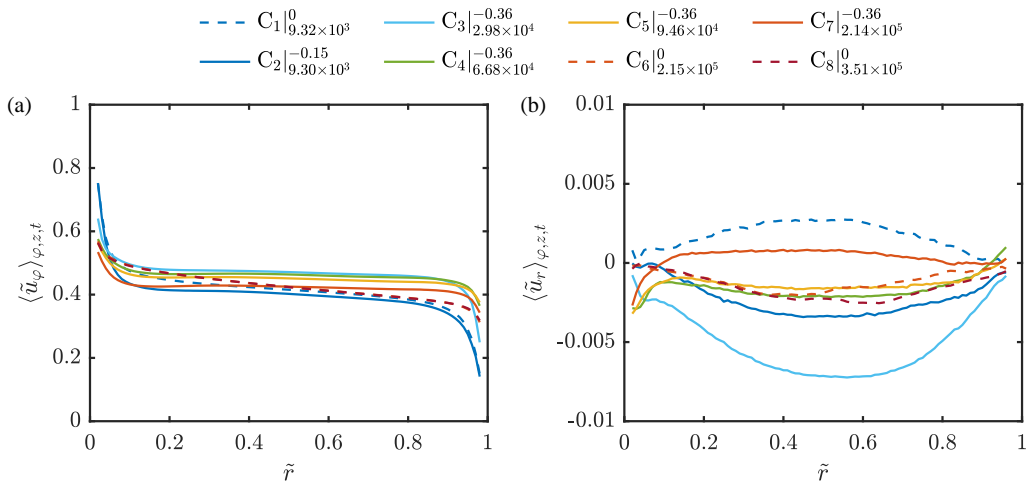}}
  \caption{Radial profiles of the (a) azimuthal and (b) radial velocity component, averaged over space (cylindrical surfaces) and time. Tilde symbols denote normalized quantities. The radial coordinate is normalized as $\tilde{r}=(r-r_1)/(r_2-r_1)$, where 0 means the location of the inner cylinder ($r_1$) and 1 the location of the outer cylinder ($r_2$). The azimuthal velocity is normalized using the cylinder speeds as $\tilde{u}_\varphi=(u_\varphi-u_{\varphi,2})/(u_{\varphi,1}-u_{\varphi,2})$ with $u_\varphi=r \omega$. To normalize the radial velocity component, the shear velocity $u_S=Re_S \nu/d$ is used: $\tilde{u}_r=u_r/u_S$. {Legend abbreviations represent $\text{C}_\# \vert _{Re_S}^{\mu}$}.}
\label{fig:mean_profs}
\end{figure}

\begin{figure}
  \centerline{\includegraphics[width=\textwidth]{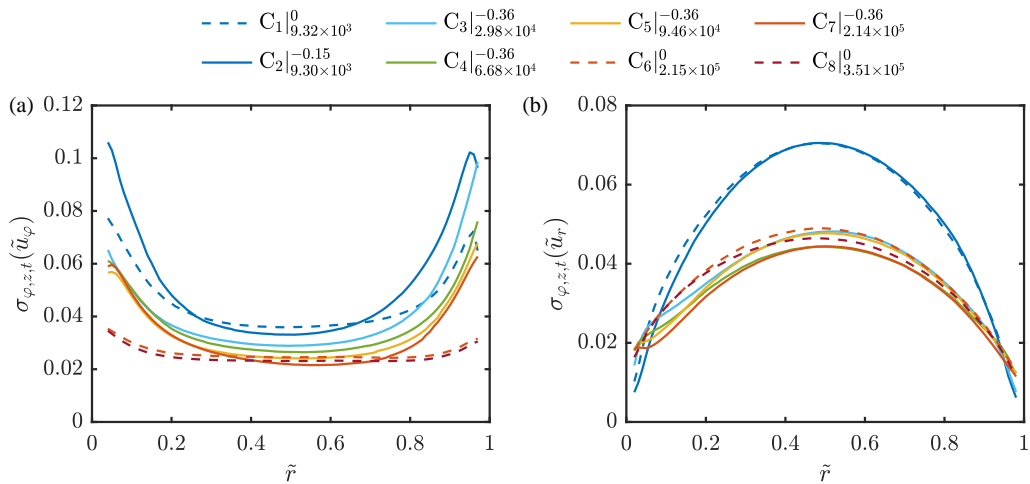}}
  \caption{Radial profiles of the standard deviation of the (a) azimuthal and (b) radial velocity component, calculated over space (cylindrical surfaces) and time. The profiles are normalized by the shear velocity $u_S$. {Legend abbreviations represent $\text{C}_\# \vert _{Re_S}^{\mu}$}.}
\label{fig:std_profs}
\end{figure}

For both flows in the classical regime (see figure \ref{fig:flow_states}a,b), large-scale rolls are visible in the {mean field} filling the whole gap. Apparently, the turbulent fluctuations {depicted in figure \ref{fig:std_profs}} are not strong enough at low shear Reynolds numbers to suppress these large-scale rolls. Further, the rolls are more pronounced at $\mu_{max}=-0.15$, indicating an increase in strength. In the ultimate regime at $\Rey_S=2.15 \times 10^5$ and $\mu=0$ {$\left( \text{C}_{6}\right)$}, the turbulent Taylor rolls disappear and nearly no axial dependence of the azimuthal velocity is visible in the mean field, just as also found numerically (see the phase diagram, figure 6 in \citet{Ostilla14b}. When the rotation rate is changed to $\mu_{max}$, Taylor vortices are formed again, which are much more pronounced {than} in the classical regime. The contour plot reveals a mushroom-like structure with distinct in- and outflow regions. The axial profile of the radial velocity component corresponding to figure \ref{fig:flow_states}d is shown in figure \ref{fig:flow_states}e. This representation exemplifies the further analysis. The recorded 23 heights capture more than one vortex pair, which is why we exclude the grayly marked data points for flow quantities calculated over the axial coordinate. The axial length of evaluation therefore starts and ends at a vortex center. In between, we use the minimum of the azimuthally-temporally averaged axial profile of the radial velocity as the location of the vortex inflow and correspondingly, the location of the vortex outflow is obtained with its maximum. The data point which is closest to a value of zero is defined as the axial location of the vortex center. In figure \ref{fig:mean_profs}, we show the normalized radial profiles of the azimuthal and radial velocity components averaged over space (cylindrical surfaces) and time ($\langle \cdot \rangle_{\varphi,z,t}$). The slopes of the profiles in the bulk nearly vanish at $\mu=\mu_{max}$, while for pure inner cylinder rotation they show a small negative slope in good agreement with other studies \citep{Ostilla14a,Brauckmann2016,Froitzheim2017}. With increasing shear Reynolds number, the difference of the data points close to the wall from the wall velocity becomes larger, which is due to the steepness of the velocity gradients. The averaged radial profiles of the radial velocity component are nearly zero all over the gap {with absolute values smaller than 1\% of the shear velocity $u_S$}, as the radial velocity {ideally} has to vanish when averaged over one vortex pair. The deviation results from the restricted axial resolution of the individual heights.

Next, the radial profiles of the standard deviation of the azimuthal and radial velocity component calculated over space (cylindrical surfaces) and time are plotted in figure \ref{fig:std_profs}. In terms of the azimuthal velocity component, the radial profiles of the standard deviation show a nearly constant low value in the center of the gap and increasing values close to the wall. This increase is more pronounced at $\mu_{max}$; while for the lowest shear Reynolds number, a peak close to the outer cylinder wall is visible. In case of the radial velocity component, the standard deviation depicts a maximum in the center of the gap and decreases towards the cylinder walls. The maximum of $\sigma_{\varphi,z,t}(\tilde{u}_r)$ for both flow cases in the classical regime is {noticeably} higher than the one in the ultimate regime. The overall shape of the profiles agrees well with the ones for pure inner cylinder rotation of \citet{Ezeta2017}, measured with PIV in the same facility at midheight.

\subsection{Probability {density function} (PDF) of velocity components}

\begin{figure}
  \centerline{\includegraphics[width=\textwidth]{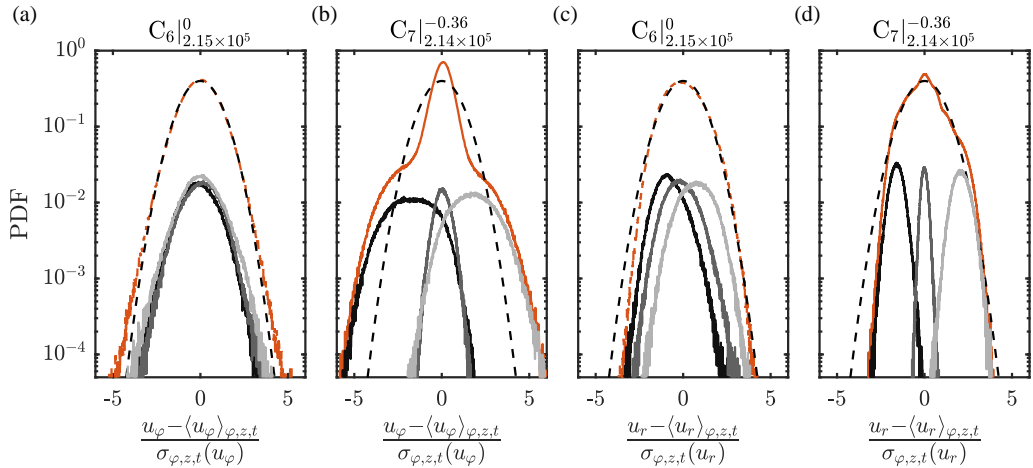}}
  \caption{Probability density functions of the (a,b) azimuthal and (c,d) radial velocity component calculated over space (cylindrical surfaces) and time at $\tilde{r}=0.5$ for $\Rey_S=2.1 \times 10^5$. {Red dashed lines correspond to $\mu=0$ $\left(\text{C}_6\right)$ and red solid lines to $\mu=-0.36$ $\left(\text{C}_7\right)$}. Dark gray, gray and bright gray lines indicate the local PDFs for the vortex inflow, vortex center and vortex outflow, respectively. The black dashed line represents a Gaussian distribution with zero mean and unit variance.}
\label{fig:PDF_uphi_ur_vgl}
\end{figure}

{To provide further validation of our measurements} and a basis for the investigation of the local PDFs of the angular momentum transport ($\S$ \ref{sec:PDF_Nu}), we first analyse the PDFs of the azimuthal and radial velocity components. In figure \ref{fig:PDF_uphi_ur_vgl}, we show the PDFs of the azimuthal and radial velocity components at $\tilde{r}=0.5$ and $Re_S=2.1 \times 10^5$ for $\mu=0$ {$\left(\text{C}_6\right)$} and $\mu=-0.36$ {$\left(\text{C}_7\right)$} as representatives. In addition, the underlying PDFs at the locations of the vortex inflow, vortex center and vortex outflow are included. 

The local PDFs at the specific vortex locations are close to Gaussian distributions, as already shown by \citet{Huisman2013} for the azimuthal velocity component measured via LDV at midheight and $\tilde{r}=0.5$. When all data at different heights are included {in} the PDFs, they still follow a Gaussian shape at $\mu=0$ for both velocity components. At $\mu_{max}$ however, the PDF of the azimuthal velocity depicts a cusp-like form centered at the origin and approximately exponential tails in accordance with the numerical simulations at a lower Reynolds number of $\Rey_S=2 \times 10^4$ by \citet{Brauckmann2016}. There and in the study of \citet{Emran2008}, a nearly identical shape was reported for the temperature PDF in RB flow, which reveals yet another clear evidence of the analogies between both systems, even on small scale statistics. In RB flow, the specific PDF shape of the temperature is induced by a combination of bursting plumes and large-scale rolls \citep{cas89,Yakhot1989,pro91}. By using the TC-RB flow analogy, we can identify here a clear fingerprint of plumes transporting angular momentum in TC flow. In addition, the vortex dominated TC flow in the region of the torque maximum in the fully turbulent regime shows similar behavior to the flow organization in RB flow. In the case of the radial velocity at $\mu_{max}$, the PDF also deviates from the ideal Gaussian shape, which can be attributed again to the intermittent bursting plumes.

\begin{figure}
  \centerline{\includegraphics[width=\textwidth]{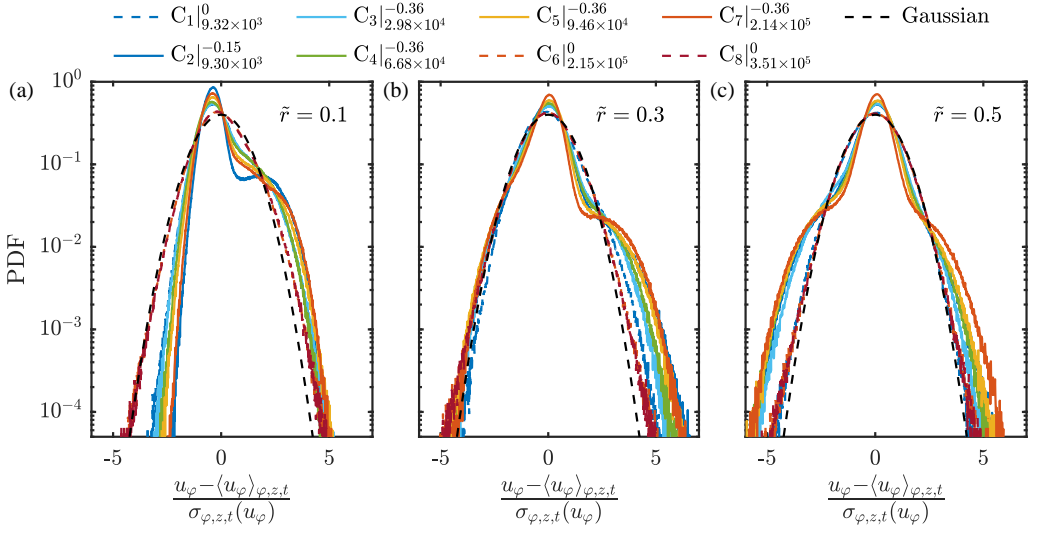}}
  \caption{Probability density functions of the azimuthal velocity component calculated over space (cylindrical surfaces) and time for the radial locations (a) $\tilde{r}=0.1$, (b) $\tilde{r}=0.3$ and (c) $\tilde{r}=0.5$. {Legend abbreviations represent $\text{C}_\# \vert _{Re_S}^{\mu}$}.}
\label{fig:PDF_uphi_r}
\end{figure}

As the PDFs of $u_r$ for flow cases at pure inner cylinder rotation depict a nearly Gaussian shape, which is also valid for different radial locations in the bulk, we omit the radial velocity component for the following analysis within this section. In figure \ref{fig:PDF_uphi_r}, we calculate the PDFs of the azimuthal velocity component for different radial locations. When the point of evaluation approaches from the gap center into the direction of the inner cylinder wall for $\mu_{max}$, the {right-hand} tail of the initial cusp-like PDF becomes more pronounced and its width increases with the shear Reynolds number. This change is caused by the dominance of the vortex outflow in the inner gap region, where the mean azimuthal velocity exhibits a large positive value. In addition, very close to the inner wall at $\tilde{r}=0.1$, both tails become increasingly exponential. This behavior is even more pronounced at higher $\Rey_S$, which is another sign of the ejection of coherent plumes from the cylinder wall. Next to these local changes, with decreasing distance to the wall, the global asymmetry of the PDFs seem to increase.

\begin{figure}
  \centerline{\includegraphics[width=\textwidth]{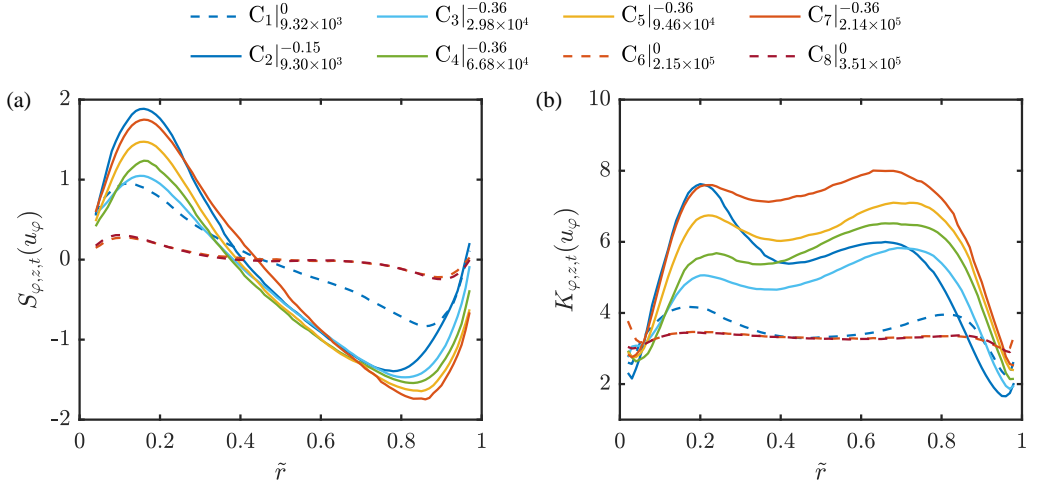}}
  \caption{Radial profiles of the (a) skewness and (b) kurtosis of the azimuthal velocity component, calculated over space (cylindrical surfaces) and time. {Legend abbreviations represent $\text{C}_\# \vert _{Re_S}^{\mu}$}.}
\label{fig:prof_skew}
\end{figure}

To account for such global properties of the PDFs, we show in figure \ref{fig:prof_skew} the radial profiles of skewness and kurtosis of the azimuthal velocity component for different Reynolds numbers. When the outer cylinder is at rest {and the ultimate regime is reached}, the skewness is close to zero and the kurtosis close to three, confirming the nearly Gaussian shape. The small radius dependent deviations in skewness for large $Re_S$ may result from remnants of turbulent Taylor vortices \citep{Lathrop92,Huisman2014,vanderveen2016b}. At $\mu_{max}$ the skewness increases from the inner cylinder wall to a maximum around $\tilde{r}=0.16$, then decreases to negative values with a minimum around $\tilde{r}=0.86$ for $\Rey_S=2.14 \times 10^5$ {$\left(\text{C}_7\right)$} and than increases again in the direction of the outer cylinder. Furthermore, the absolute skewness increases with the shear Reynolds number. This behavior is similar to the one of the temperature {fields} in RB flows reported by \citet{Emran2008}.
The kurtosis profiles at $\mu_{max}$ depict two maxima, one in the inner gap region at $\tilde{r}=0.22$ and one in the outer gap region at $\tilde{r}=0.68$ for the highest $\Rey_S$. This reflects that the non-Gaussianity of the PDFs is most pronounced in the regions dominated by the vortex in- and outflows due to coherent plumes. {In summary, the global and local velocity field statistics of our measurements agree very well with those of the mentioned literature and exceed the state of the art especially for $\mu_{max}$ to higher forcings.}

\section{Angular momentum transport} 
{As the angular momentum transport, expressed in terms of the quasi-Nusselt number $\text{Nu}_\omega$, is strongly influenced by the local and global flow organization inside the gap, it is an appropriate parameter to statistically investigate the existence of flow structures. Therefore, within this section, we first analyze the global angular momentum transport to compare its amount with the literature, and second, we analyze the axial dependent radial profiles of $\text{Nu}_\omega$ to reveal the most relevant axial locations for the transport. The subsequent results are fundamental for the small-scale statistical analysis of the local momentum transport in the next section and enable new insights into the vortex-dominated momentum transport.}
The Nusselt number is composed of a convective and a viscous part \citep{Eckhardt07b}:

\begin{equation} \label{Nu_total}
\text{Nu}_\omega=\text{Nu}_\omega^c (r) + \text{Nu}_\omega^\nu (r)=\frac{r^2}{J_{lam}} \left\langle u_\varphi u_r \right\rangle_{\varphi,z,t} -\frac{\nu r^3}{J_{lam}}\partial_r \left\langle \frac{u_\varphi}{r} \right\rangle_{\varphi,z,t}.
\end{equation}

\noindent While the viscous term $\text{Nu}_\omega^ \nu$ dominates in the boundary layers, the convective term $\text{Nu}_\omega^c$ dominates in the bulk. Since the focus of our investigation is set to the bulk region, we neglect thus the viscous part. Furthermore, as it is shown in figure \ref{fig:mean_profs}b, the radial velocity component nearly vanishes when averaged over cylindrical surfaces with an axial length of one vortex pair: {$\langle u_r \rangle_{\varphi,z,t} \approx 0$. Therefore, $r^2\langle u_r u_\varphi \rangle_{\varphi,z,t} \approx r^2\langle u_r^\prime u_\varphi^\prime \rangle_{\varphi,z,t}$ is valid, which means that only the fluctuations of the azimuthal velocity component around its mean profile contribute to the net momentum flux through these cylindrical surfaces \citep{Brauckmann2016}}. Accordingly, the net convective flux in the bulk flow can be calculated as

\begin{align}
\text{Nu}_\omega^{c,net} &= \frac{r^2}{J_{lam}} \left\langle u_\varphi^\prime u_r^\prime \right\rangle_{\varphi,z,t}, \quad \textrm{with} \\
u_r^\prime (r,\varphi,z,t) & =u_r (r,\varphi,z,t)-\langle u_r (r,\varphi,z,t) \rangle_{\varphi,z,t}, \nonumber \\ 
u_\varphi^\prime (r,\varphi,z,t) & =u_\varphi (r,\varphi,z,t)-\langle u_\varphi (r,\varphi,z,t) \rangle_{\varphi,z,t}. \nonumber
\end{align}

\begin{figure}
  \centerline{\includegraphics[width=\textwidth]{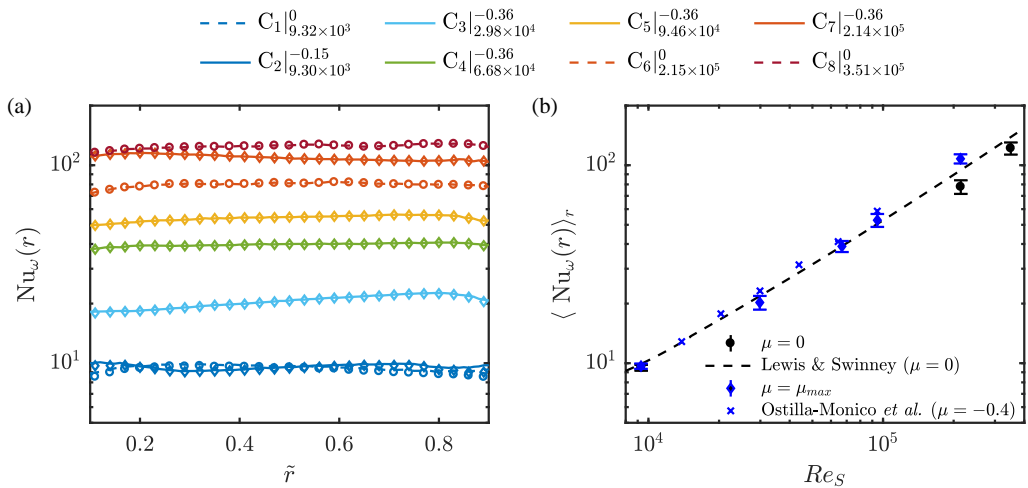}}
  \caption{(a) Azimuthal and time averaged profiles of the total Nusselt number indicated by lines and of the net convective Nusselt number indicated by diamonds ($\mu_{max}$) and circles ($\mu=0$). The evaluation is restricted to the interval $0.1 \leq \tilde{r} \leq 0.9$ {to exclude the boundary layers and focus on the bulk flow}. {Legend abbreviations represent $\text{C}_\# \vert _{Re_S}^{\mu}$}. (b) Radially averaged total Nusselt number as function of $\Rey_S$. Errorbars represent the standard deviation of the total Nusselt number along the radial coordinate. Data are compared to torque measurements of \citet{Lewis99} ($\mu=0$, $\eta=0.724$) and DNS of \citet{Ostilla14b} ($\mu=-0.4$, $\eta=0.714$).}
\label{fig:prof_Nusselt}
\end{figure}

To avoid confusion, in the following, we will call the Nusselt number $\text{Nu}_\omega=\text{Nu}_\omega^c + \text{Nu}_\omega^\nu $ shown in equation \ref{Nu_total}, the \textit{total} Nusselt number. In figure \ref{fig:prof_Nusselt}a, we show the radial profiles of $\text{Nu}_\omega$ indicated by lines, and the net convective momentum flux $\text{Nu}_\omega^{c,net}$, indicated by diamonds ($\mu_{max}$) and circles ($\mu=0$). The analysis is restricted to the bulk region with $\tilde{r} \in [0.1,0.9]$. {A slight dependence of the $\text{Nu}_\omega$-profiles on the radial coordinate is observable, which differs partially for the different flow states. The deviation from the predicted conservation of the momentum transport along the radial direction according to \citet{Eckhardt07b} is probably due to the finite axial length of our experimental setup in contrast to their theory, which is based on cylinders of infinite length. However, the effect of the cylinder length is reduced in our setup by cutting the axial length of evaluation to the size of one vortex pair. Moreover, also the vortex aspect ratio influences the value of $\text{Nu}_\omega$ as a function of $Re_S$ \citep{Huisman2014,Martinez2014}, which takes values for the current study in the range of $2.13d-2.67d$. Considering these aspects, the shape of the $\text{Nu}_\omega$-profiles is satisfactory.} The net convective momentum flux dominates the total Nusselt number and is valid to evaluate the momentum transport in the bulk region \citep{Huisman2012}. Furthermore, in figure \ref{fig:prof_Nusselt}b, we show the radially averaged total Nusselt number as a function of $\Rey_S$, which is also compared to both torque measurements of \citet{Lewis99} ($\mu=0, \eta=0.724$) and DNS of \citet{Ostilla14b} ($\mu=-0.4, \eta=0.714$). We find a very good agreement with these data, which enables a more detailed analysis of the momentum transport. 

\begin{figure}
  \centerline{\includegraphics[width=\textwidth]{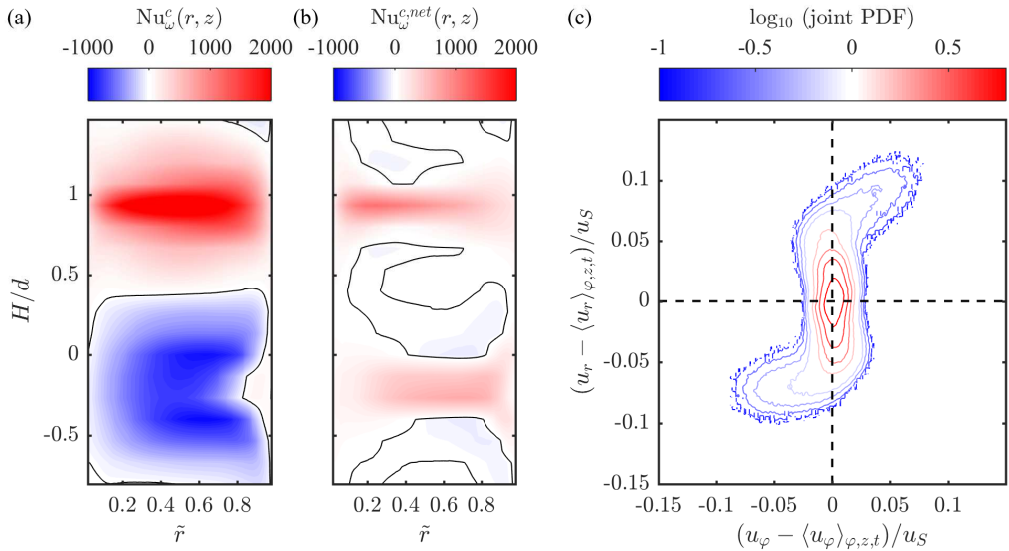}}
  \caption{Contour plot of the dimensionless (a) convective and (b) net convective {local} angular momentum transport in a meridonal plane for $Re_S=2.14 \times 10^5$ and $\mu=-0.36$ {($\text{C}_7$)}. Black lines indicate the zero line of the depicted quantities. The color codes are given in the legends. (c) Joint PDF of the radial and azimuthal velocity fluctuations for the same flow state at $\tilde{r}=0.5$. The colors give the probability (in log-scale), see legend.}
\label{fig:Nu_contours}
\end{figure}

In order to get deeper insight into the relation between the convective and net convective Nusselt number, we plot in figures \ref{fig:Nu_contours}a,b both local quantities in a meridional plane ($r-z$ plane) for $Re_S=2.14 \times 10^4$ and $\mu=-0.36$ {$\left(\text{C}_7\right)$}. The local Nusselt number can be much larger than its average value as already noticed by \citet{Huisman2012} and \citet{Ostilla14c}. In the case of the convective Nusselt number $\text{Nu}_\omega^c$ (figure \ref{fig:Nu_contours}a), the momentum transport in the area of the vortex outflow is positive and strongly concentrated to a small axial region, while in the area of the vortex inflow, the transport is negative and less focused. The magnitude of positive momentum flux is twice as large as the negative one, which yields in average a positive net transport. Alternatively,  when the net transport is plotted (figure \ref{fig:Nu_contours}b), a much clearer picture of the transport process is revealed. While in the in- and outflow region the net transport is positive, only in the sheared regions in between negative net transport can be detected. In order to explain this difference, in figure \ref{fig:Nu_contours}c we show the joint PDF of the radial and azimuthal velocity fluctuations at $\tilde{r}=0.5$. In the inflow region, where {the} fluid is transported strongly in the negative radial direction, the azimuthal velocity depicts a high probability for negative fluctuation values. As a consequence, the net transport has to be positive. In the outflow region, both velocities are mainly positive, resulting also in a positive correlation. However, when the mean azimuthal velocity component is not subtracted, the joint PDF is shifted to the right, leading to negative correlations in the inflow region (not shown). In summary, the difference between $\text{Nu}_\omega^{c}$ and $\text{Nu}_\omega^{c,net}$ is the transport of the mean azimuthal velocity by the turbulent Taylor vortices, which vanishes when averaged over cylindrical surfaces. By neglecting this fraction, a much clearer picture of the transport process is revealed. In addition, the representation of the Nusselt number in the meridional plane demonstrates the importance of the in- and outflow regions for the net convective transport. 

\begin{figure}
  \centerline{\includegraphics[width=\textwidth]{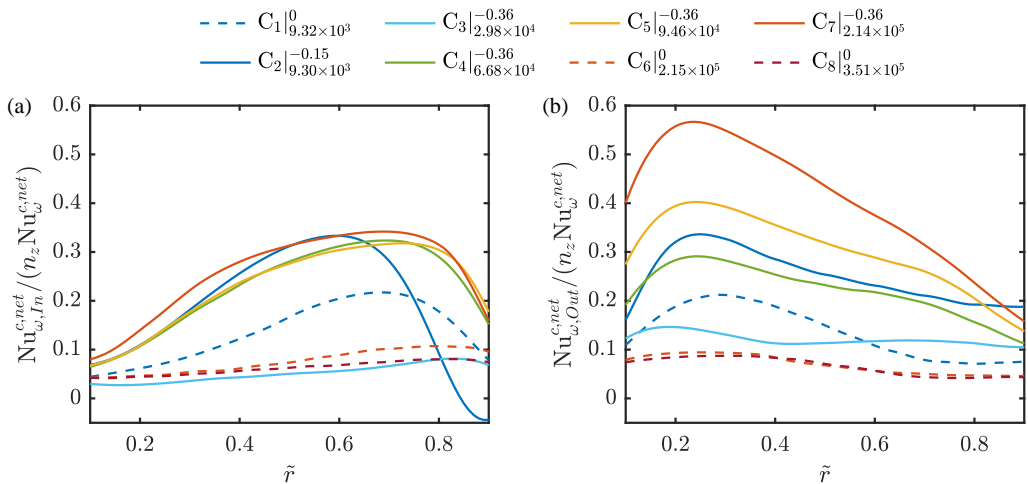}}
  \caption{Radial profiles of the contribution of the net convective momentum transport of the (a) inflow and (b) outflow to the total transport. $n_z$ represents the number of heights included into the average over cylindrical surfaces. {Legend abbreviations represent $\text{C}_\# \vert _{Re_S}^{\mu}$}.}
\label{fig:Nu_vortex_contribution}
\end{figure}

The contribution of the vortex in- and outflow as {a} function of the radial location and $\Rey_S$ is depicted in figure \ref{fig:Nu_vortex_contribution}. As a global feature, the contribution of the vortex inflow to the total net convective momentum transport is especially pronounced in the outer gap region ($\tilde{r}>0.5$) while the opposite is true for the outflow. For the two flow states in the classical regime, where the values of the total Nusselt numbers are comparable (see figure \ref{fig:prof_Nusselt}), the contributions of the in- and outflow are much more dominant at $\mu_{max}$. This reflects the strong correlation of the enhanced momentum flux at $\mu_{max}$ and the turbulent Taylor vortices. Further, in the ultimate regime at $\mu=0$, again the effect of remnants of these large vortices becomes visible in the slight dependence of the Nusselt number on $\tilde{r}$. When the ultimate regime is reached at $Re_S=2.98\times 10^4$ at $\mu_{max}$ {$\left(\text{C}_3\right)$}, the contributions of the vortex in- and outflow are small and become significant again at $Re_S=6.68 \times 10^4$ {$\left(\text{C}_4\right)$}. For even higher shear Reynolds numbers, the contribution of the inflow stays nearly constant with a fraction slightly above 30\% in the outer gap region; while the outflow contribution continuously increases up to approximately 60\% in the inner gap region. This value is strikingly high and demonstrates that at very high Reynolds numbers, the net convective transport shrinks to very small axial regions, where most of the momentum transport takes place. {We would like to encourage further numerical or experimental studies, to confirm this finding.}

\section{Probability density function of the net convective angular momentum flux} \label{sec:PDF_Nu}
 
With the knowledge of the total net convective transport and the relative contributions of the in- and outflow, we analyze now the PDFs of $\text{Nu}_\omega^{c,net}$ {to identify statistical footprints of small-scale plume structures.} {To properly normalize these PDFs and provide an idealized shape for comparison, we firstly introduce the Gaussian distribution.} According to figure \ref{fig:PDF_uphi_ur_vgl}, the PDFs of the radial and azimuthal velocity components are nearly Gaussian for pure inner cylinder rotation. In that case, the PDF of their product can be described according to \citet{Thoroddsen1992} and \citet{Chu1996} based on a Gaussian distribution. The PDF of two jointly Gaussian random variables $x$ and $y$ is given by

\begin{equation} \label{equ:joint_PDF_ell}
P(x,y)=\frac{1}{\pi \sigma_x \sigma_y \sqrt{1-\rho_P^2}} \exp \left[ \frac{1}{2(1-\rho_P^2)} \left(\frac{x^2}{\sigma_x^2}-\frac{2xy\rho_P}{\sigma_x \sigma_y}+\frac{y^2}{\sigma_y^2} \right) \right],
\end{equation} 

\noindent with the individual standard deviations $\sigma_x$ and $\sigma_y$ and the correlation coefficient ${\rho_P=\langle xy \rangle/(\sigma_x \sigma_y)}$. Equation (\ref{equ:joint_PDF_ell}) represents an inclined elliptic shape for the joint PDF in contrast to the one shown in figure \ref{fig:Nu_contours}c. Furthermore, the PDF of the product $z=xy$ is \citep{Thoroddsen1992,Chu1996}

\begin{equation} \label{PDF_joint_gaussian}
 P(z)=\frac{1}{\pi \sigma_x \sigma_y \sqrt{1-\rho_P^2}} \exp \left( \frac{\rho_P z}{\left(1-\rho_P^2 \right) \sigma_x \sigma_y} \right) K_0 \left(\frac{\vert z \vert}{\left(1-\rho_P^2 \right) \sigma_x \sigma_y} \right),
 \end{equation}

\noindent with $K_0$ the modified Bessel function of the second kind. The prefactor of the exponential function in equation (\ref{PDF_joint_gaussian}) {is} used to normalize the different PDFs of the angular momentum flux, i.e to standardize the width of their tails. It is worth mentioning that the correlation coefficient $\rho_P$ cannot be normalized in a proper way, which leads to different slopes of the exponential tails of the prediction depending on $\rho_P$. Therefore, in our calculation,  we use the prediction according to equation (\ref{PDF_joint_gaussian}) for the case {$Re_S=3.51\times10^5$ and $\mu=0$ ($\text{C}_8$)}. 

\begin{figure}
  \centerline{\includegraphics[width=\textwidth]{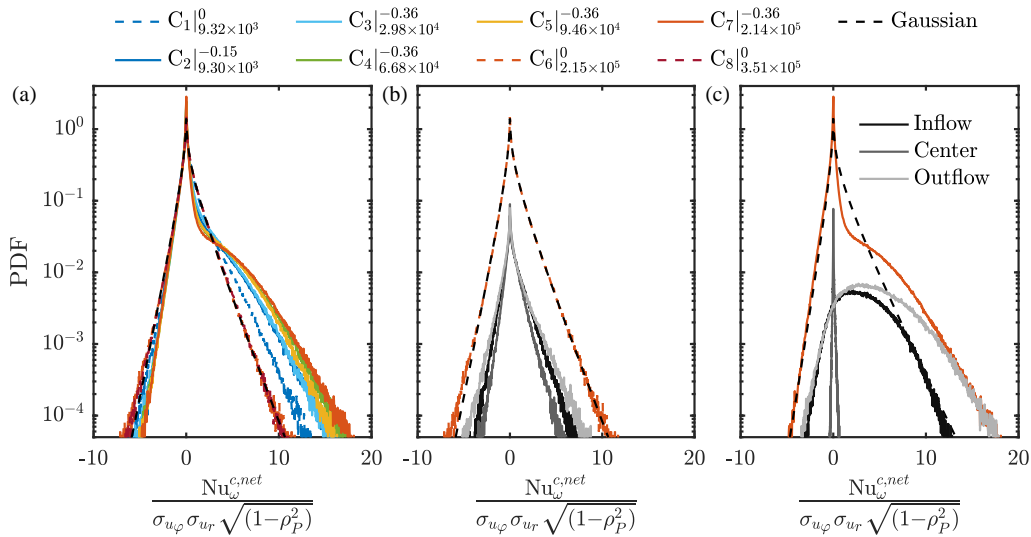}}
  \caption{Probability density functions of the local net convective angular momentum transport calculated over space (cylindrical surfaces) and time for $\tilde{r}=0.5$. (a) PDFs of all investigated flow states are depicted. (b) PDF for $Re_S=2.15\times10^5$ and $\mu=0$ {($\text{C}_6$)} with the corresponding PDF at the axial height of vortex inflow, vortex center and vortex outflow. (c) Same plot as in (b) for $Re_S=2.14\times10^5$ and $\mu=-0.36$ {($\text{C}_7$)}. The dashed black line indicates the prediction according to equation (\ref{PDF_joint_gaussian}). {Legend abbreviations represent $\text{C}_\# \vert _{Re_S}^{\mu}$}.}
\label{fig:PDF_Nuc1}
\end{figure}

In figure \ref{fig:PDF_Nuc1}, we show the PDFs of $\text{Nu}_\omega^{c,net}$, which are normalized with the factor $\sigma_{u_r} \sigma_{u_\theta} \sqrt{1-\rho_P^2}$ in order to compare their shapes, at $\tilde{r}=0.5$. For pure inner cylinder rotation, the PDFs of the net convective momentum transport agree very well with the proposed jointly Gaussian prediction. However, at $\mu_{max}$ the PDFs become highly skewed due to a change of shape in the positive tails. This deviation results in an increasing but still rare number of positive extreme events of momentum flux, which becomes more pronounced with increasing $\Rey_S$. According to figure \ref{fig:PDF_Nuc1}c, these rare and extreme events can be almost completely attributed to the vortex in- and especially the vortex outflow {due to changes in the azimuthal velocity component (see also figure \ref{fig:PDF_uphi_ur_vgl})}. Here, a strong correlation between the radial and azimuthal velocity component exists due to the coherence of the plumes (see also figure \ref{fig:Nu_contours}c). In addition, all PDFs have in common that {zero is the value with the highest probability} instead of the average value of the Nusselt number. The overall shape of the PDFs at $\mu_{max}$ is in good agreement with the findings of \citet{Brauckmann2016b}, {whose analysis was however restricted to low Reynolds numbers ($\Rey_S=2\times 10^4$) and PDFs along cylinder surfaces}. Furthermore, our PDFs are comparable to the corresponding PDFs of the heat flux in {the} RB flow. \citet{Shang2004} reported similar PDF shapes in the case of RB convection. They argue that such large rare events are footprints of heat flux fluctuations induced by thermal plumes. This feature is obviously shared with TC flows and supports the origin of large-scale turbulent vortices as the result of small-scale unmixed plumes. {Even more, slight counter-rotating cylinders at $\mu_{max}$ instead of pure inner cylinder rotation seems to be the right kinematic boundary condition of TC flows for comparisons with the RB flow concerning $\text{Nu}_\omega$-PDFs.}

\begin{figure}
  \centerline{\includegraphics[width=\textwidth]{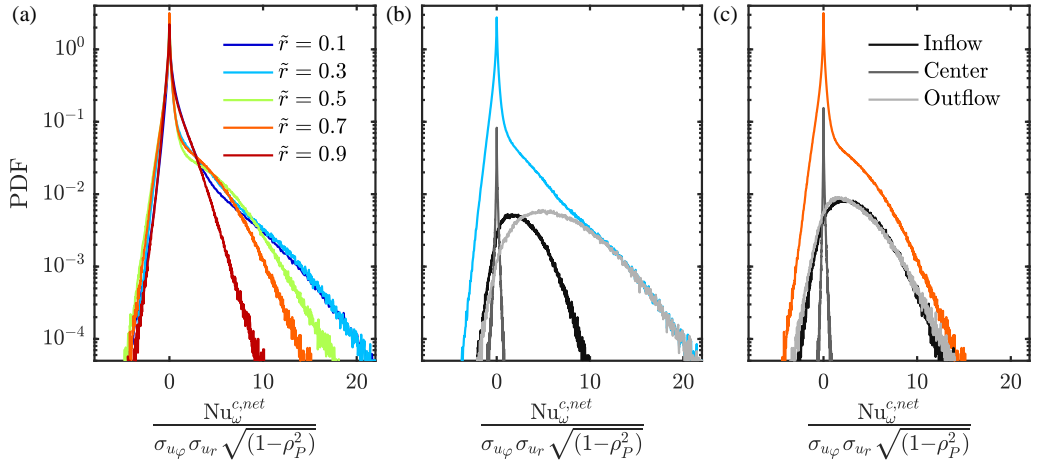}}
  \caption{(a) Probability density functions of the local net convective angular momentum transport calculated over space (cylindrical surfaces) and time for $Re_S=2.14 \times 10^5$ at $\mu=-0.36$ at different radial positions {($\text{C}_7$)}. (b,c) Same PDFs at $\tilde{r}=0.3$ and $\tilde{r}=0.7$, respectively with the corresponding PDFs related to the vortex inflow, vortex center, and vortex outflow.}
\label{fig:PDF_Nuc2}
\end{figure}

In figure \ref{fig:PDF_Nuc2}, we show the PDFs of $\text{Nu}_\omega^{c,net}$ for various different radii $r$ for the case of $Re_S=2.14 \times 10^5$ at $\mu=-0.36$ {($\text{C}_7$)}. The negative tails of the PDFs are largely identical for all investigated radial positions, while the right tails strongly depend on $\tilde{r}$. From the inner to the outer cylinder wall, the width of the positive tail and therefore the asymmetry decreases {coinciding with degeneration of the exponential tails in the $u_\varphi$-PDFs in figure \ref{fig:PDF_uphi_r}}. As was shown in figure \ref{fig:Nu_vortex_contribution}b, the maximum of the outflow contribution to the overall momentum transport is close to $\tilde{r}=0.3$. At this location, the right tail of the PDF reflects extreme positive events due to the emission of plumes and is nearly covered by the PDF of the outflow (see figure \ref{fig:PDF_Nuc2}b). However, in the outer gap region at $\tilde{r}=0.7$, the asymmetric right tail of the PDF is comparably formed by both the in- and outflow region. {From our point of view, our findings demonstrate that the comparative evaluation of axially global and local PDFs of $\text{Nu}_\omega^{c,net}$, which has not been done in TC flow before, is crucial for deeper insights into the statistics of the momentum transport.}

\section{Azimuthal energy co-spectra and correlations}
\subsection{Azimuthal energy co-spectra} \label{sec:Azimuthal energy co-spectrum}

As shown in the previous section, the net convective Nusselt numbers suggest the presence of small-scale plumes concentrated {especially} in the in- and outflow regions of the turbulent Taylor vortices, which dominate the transport. Hence, we analyze the spatial energy co-spectra to detect the presence {and lengthscale} of small-scale structures {in the gap}. We assume velocity fluctuations at a constant radial $r_c$ and axial $z_c$ position in the homogeneous $\varphi$-direction with a total number of azimuthal points of $n_\varphi=0,1,...,N-1$ equidistantly spaced by $\Delta s = \Delta \varphi r_c$ i.e.

\begin{align}
 		u_r^* \left(r_c,\varphi,z_c,t \right) &=u_r\left(r_c,\varphi,z_c,t \right)- \langle u_r \left(r_c,\varphi,z_c,t \right) \rangle _t, \\
		u_\varphi^* \left(r_c,\varphi,z_c,t \right) &=u_\varphi \left(r_c,\varphi,z_c,t \right)- \langle u_\varphi \left(r_c,\varphi,z_c,t \right) \rangle _t.
\end{align}
 
\noindent The discrete spatial Fourier transform $U_{r,\varphi}$ of both fluctuation components $u^*_{r,\varphi}$ is given by

\begin{equation}
	U_{r,\varphi} \left(n_\varphi \right) = \sum_{k=0}^{N-1} u_{r,\varphi}^* \left( k \right) \exp \left(- \dfrac{2 \pi i k n_\varphi}{N} \right),
\end{equation}

\noindent where for simplicity we have not written out the dependences on $r_c$, $z_c$ and $t$. Thus, the spatial energy co-spectrum $E_{r \varphi}$ can be calculated as

\begin{equation}
E_{r \varphi} \left( k_{\varphi}^{n_\varphi} \right)= 
\begin{cases} \dfrac{1}{N^2} \vert U_{r}^{n_\varphi} \cdot U_{\varphi}^{n_\varphi} \vert, & \mbox{for } n_\varphi=[0,\frac{N}{2}] \\[0.8em] \dfrac{1}{N^2} \left( \vert U_{r}^{n_\varphi} \cdot U_{\varphi}^{n_\varphi} \vert + \vert U_{r}^{N-n_\varphi} \cdot U_{\varphi}^{N-n_\varphi} \vert \right), & \mbox{for } n_\varphi =[ 1,...,\frac{N}{2}-1] ,
\end{cases}
\end{equation}
 
\noindent with the wavenumber vector $k_\varphi^{n_\varphi}=\left(\Delta s \right)^{-1} n_\varphi /N$. The co-spectra are determined for each time step $t$ and afterwards ensemble-averaged over 1500 snapshots for every case. To enable a comparison of the co-spectra for different $\Rey_S$ and radial positions, we normalize all co-spectra with the area under its graph ${A_{E} \approx (2 \Delta s)^{-1}\sum_{n_{\varphi}=0}^{N/2-1} \left[ E_{r \varphi}(k_\varphi^{n_\varphi}) + E_{r \varphi}(k_\varphi^{n_\varphi+1}) \right]}$ based on the trapezoidal integration method.
 
\begin{figure}
  \centerline{\includegraphics[width=\textwidth]{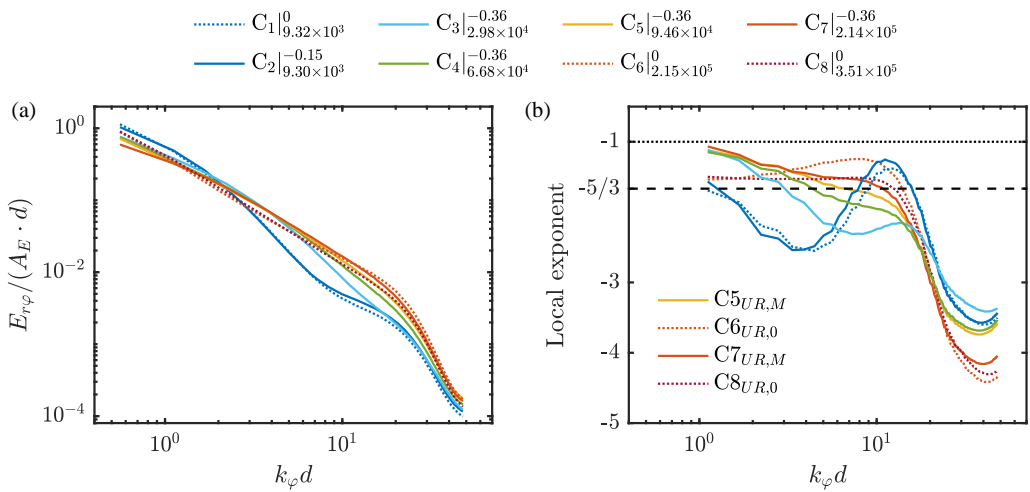}}
  \caption{(a) Temporally and azimuthally averaged azimuthal kinetic energy co-spectra evaluated at $\tilde{r}=0.5$. Spectra are normalized to cover an area of 1 under their respective curves and by the gap width d. (b) Local scaling exponent $\gamma$ of the co-spectra for $E_{r,\varphi} \sim k_\varphi ^\gamma$ calculated with a bin size of of $\log_{10}(k_\varphi)=0.5$. {Legend abbreviations represent $\text{C}_\# \vert _{Re_S}^{\mu}$}.}
\label{fig:spectra_scale}
\end{figure}

First of all, we show the temporally and axially averaged azimuthal energy co-spectra at $\tilde{r}=0.5$ in figure \ref{fig:spectra_scale}a {to illustrate the scaling of our spectra and compare it with other studies}. The kinetic energy co-spectra show that most of the energy lies within the large scales, corresponding to small wavenumbers. The co-spectra depict a noticeable drop for $k_\varphi d \approx 20$. It is worth mentioning that we do not see any peak in the large-scale regime, because of the limited range of the azimuthal coordinate in the experiments. Moreover, we cannot resolve the energy content of axisymmetric Taylor rolls, as they correspond to an azimuthal wavenumber of $k_\varphi=0$.

In the case of $Re_S$ being in the classical regime, the spectra show, compared to the other flow states, a stronger decrease of the spectral energy at {mid scales} and a kink in the region of $10 \leq k_\varphi d \leq 20$. When $\Rey_S$ is increased into the ultimate regime (with $\mu$ fixed at $\mu=\mu_{max}$), this kink continuously diminishes and energy is redistributed from the large to the small scales. Based on the power-law ansatz $E_{r,\varphi} \sim k_\varphi ^\gamma$, the local scaling exponent $\gamma$ corresponding to the co-spectra is shown in figure \ref{fig:spectra_scale}b. $\gamma$ is calculated with a bin size of $\log_{10}(k_\varphi)=0.5$. {It is worth {mentioning} that $\gamma$ is sensitive to the data processing and the accompanying error propagation}. {$\gamma$ is given here for this choice of bin size and to show the trend of $\gamma$ with d.} We observe that the spectra neither show $-1$ nor $-5/3$ scaling, which is consistent with the results of \citet{Lewis99}, \citet{Ostilla2016} and \citet{Huisman2013}. In the ultimate regime (with $\mu$ fixed at $\mu=\mu_{max}$), the exponent decreases with increasing $k_\varphi$ before it strongly drops down in the viscous regime beyond $k_\varphi d \approx 20$. This decrease becomes smaller with increasing $\Rey_S$. 

However, in the case of the highest investigated shear Reynolds number at $Re_S=3.51 \times 10^5$ and $\mu=0$ {($\text{C}_8$)}, the local exponent is nearly constant for $k_\varphi d \leq 10$ with a value of $\gamma \approx -1.52$, which is slightly above $-5/3$.

\begin{figure}
  \centerline{\includegraphics[width=\textwidth]{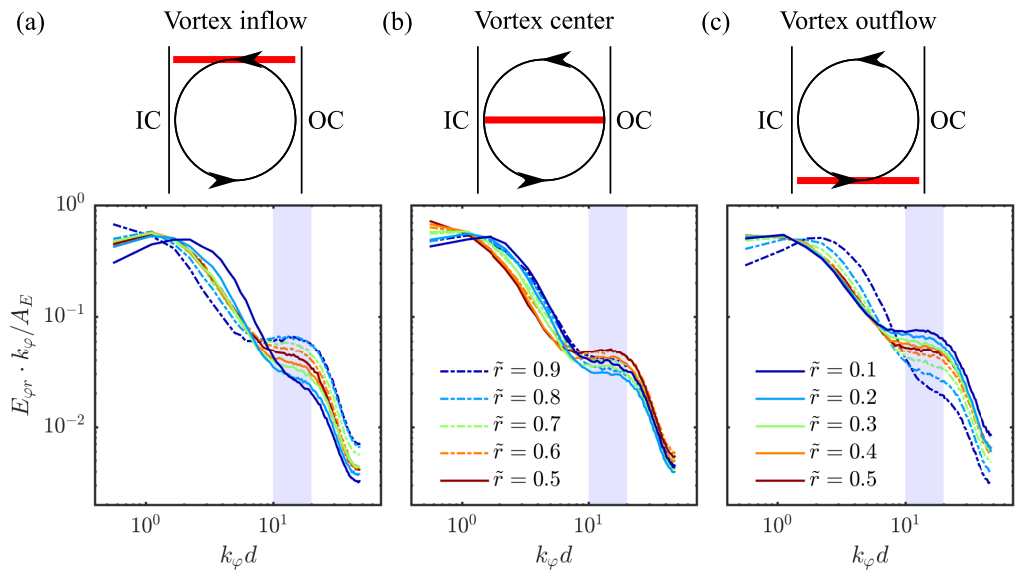}}
  \caption{Temporally averaged pre-multiplied azimuthal kinetic energy co-spectra in the classical regime for {$Re_S=9.30\times10^3$ and $\mu=-0.15$ ($\text{C}_2$)} at different radial positions at the axial height of (a) vortex inflow, (b) vortex center and (c) vortex outflow. The region of $k_\varphi d \in [10,20]$ with enhanced plume emission is marked in light blue. Sketches above the co-spectra for the different vortex regions are added for clarity.}
\label{fig:spectra_CR}
\end{figure}

Next, we focus on the local pre-multiplied energy co-spectra at the axial height of the vortex inflow, vortex center, and vortex outflow respectively, {where the PDF analysis of $\text{Nu}_\omega^{c,net}$ suggests the occurrence of small-scale intermittent plumes.} \textit{Pre-multiplied} means that the co-spectra are multiplied with the wavenumber vector $k_\varphi$, such that the area under its graph corresponds to the kinetic energy \citep{Smits2011}. In figure \ref{fig:spectra_CR}, we show the pre-multiplied energy co-spectra in the classical regime for $\Rey_S=9.32 \times 10^3$ and $\mu=-0.15$ {($\text{C}_2$)} at the three vortex positions. Firstly, we address the peak that corresponds to the large scales. For the vortex inflow in figure \ref{fig:spectra_CR}a, {the large-scale peak is located around $k_\varphi d \approx 1.1$ within the bulk for $0.2\leq \tilde{r}\leq 0.8$, while close to the outer cylinder at $\tilde{r}=0.9$, the peak lies outside of our resolvable scales}. Close to the IC wall, the peak shifts to $k_\varphi d \approx 2.2$ at $\tilde{r}=0.1$. In figure \ref{fig:spectra_CR}c, at the location of the vortex outflow, the opposite behavior is observed. Here, the large-scale peak shifts to smaller wavenumbers when the radial position increases from the IC to the OC. At the vortex center in figure \ref{fig:spectra_CR}b, the large-scale peak is shifted from smaller wavenumbers in the center of the gap ($\tilde{r}=0.5$) to larger ones at both cylinder walls, in an almost symmetric manner. This suggests that the formation of this large-scale peak is connected to the mean radial velocity field, which is in itself caused by the large-scale turbulent Taylor rolls. When fluid impacts on the cylinder walls due to these rolls, structures of the size $k_\varphi d \approx 2.2$ are formed. Since the axisymmetric flow has a wavenumber of $k_\varphi=0$, the existence of a large-scale peak may be an indication of modified turbulent Taylor vortices. This point will be discussed in more detail in $\S$ \ref{sec:CPOD}.

With respect to the small-scales peak, we observe that it is located at wavenumbers around $k_\varphi d \in [10,20]$ for all three depicted heights, independent of the radial coordinate. However, its amplitude strongly varies with $\tilde{r}$ and the height $z$. In the region of the vortex inflow shown in figure \ref{fig:spectra_CR}a, the small-scale peak is most pronounced near the OC and its amplitude decreases monotonically with decreasing $\tilde{r}$. On the contrary, at the height of the vortex outflow in figure \ref{fig:spectra_CR}c, the small-scale peak amplitude is largest close to the IC and decreases in amplitude towards the OC. In the region of the vortex center in figure \ref{fig:spectra_CR}c, we find the highest amplitude of the small-scale peak in the center of the gap. This is due to the emission of coherent plumes from the cylinder walls which give rise to the formation of Taylor rolls, as was already mentioned before. Thus, this peak should indeed be most pronounced in the ejecting regions, i.e. in the outflow region at the IC and in the inflow region at the OC, respectively. At the height of the vortex center ---where no predominant flow direction concerning the radial velocity component is present--- the behavior is different: Plumes rise from both cylinder walls and travel towards the bulk flow which results in the highest peak amplitude at $\tilde{r}=0.5$. {Considering the fact, that the contribution of the vortex center to extreme and strong events of momentum flux is almost negligible (see figures \ref{fig:PDF_Nuc1}, \ref{fig:PDF_Nuc2}), these detected plumes seem to compensate there total radial momentum transport.}

\begin{figure}
  \centerline{\includegraphics[width=\textwidth]{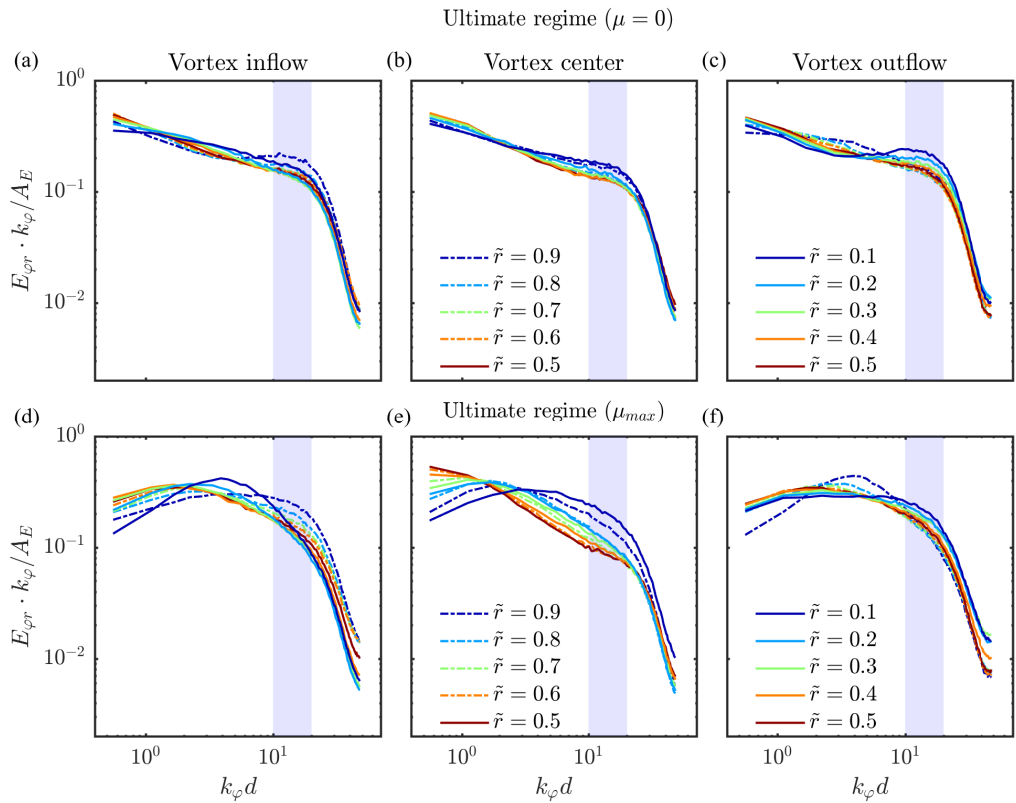}}
  \caption{Upper row: Temporally averaged pre-multiplied azimuthal kinetic energy co-spectra for {$Re_S=2.15\times10^5$ and $\mu=0$ ($\text{C}_6$}) at different radial positions at the axial height of (a) vortex inflow, (b) vortex center and (c) vortex outflow. Region of $k_\varphi d \in [10,20]$ is marked in light blue. Lower row: Same spectra for {$Re_S=2.14\times10^5$ and $\mu=-0.36$ ($\text{C}_7$}) at the axial height of (d) vortex inflow, (e) vortex center and (f) vortex outflow.}
\label{fig:spectra_UR}
\end{figure}

The pre-multiplied energy co-spectra in the ultimate regime at $\Rey_S=2.1 \times 10^5$ for the three vortex locations are depicted in figure \ref{fig:spectra_UR} for $\mu=0$ ({$\text{C}_6$},a-c) and $\mu_{max}$ ({$\text{C}_7$},d-f). For $\mu=0$, no large-scale peak can be seen in any of the co-spectra at the investigated heights. This is consistent with the finding shown in figure \ref{fig:flow_states}c, where we observed that the Taylor rolls have faded away. Accordingly, the shape of the co-spectra is less dependent on both the axial coordinate and on $\tilde{r}$. However, the co-spectra show a prominent change in the slope around $k_\varphi d \approx 20$: In figure \ref{fig:spectra_UR}a (vortex inflow), a small-scale peak is formed around $k_\varphi d \in [10,20]$ close to the OC at $\tilde{r}=0.9$. This is comparable to the previous case in the classical regime. Also in the region of the vortex outflow (figure \ref{fig:spectra_UR}c), we observe a peak within the same range of scales at $\tilde{r}=0.1$. For the vortex center however (figure \ref{fig:spectra_UR}b), no peak is visible.

Also when $\mu$ is changed to $\mu_{max}$ (see figures \ref{fig:spectra_UR}d-e), we identify a similar behavior as in the classical regime, although the energy is more homogeneously distributed over all scales. At the height of the vortex inflow as seen in figure \ref{fig:spectra_UR}d, a large-scale peak is present which shifts to $k_\varphi d \approx 3.9$ at $\tilde{r}=0.1$. At the vortex outflow in figure \ref{fig:spectra_UR}f, this shift appears close to the OC wall at $\tilde{r}=0.9$ for the same wavenumber. In the region of the vortex center (figure \ref{fig:spectra_UR}e), the large-scale peak is most pronounced at $\tilde{r}=0.5$ at a smaller wavenumber. For all the heights explored, the co-spectra do not reveal a peak in the small-scale regime. However, at the vortex inflow the energy is redistributed continuously from large to small scales when $\tilde{r}$ is varied from 0.1 to 0.9, as it is clearly visible in the marked regime of $k_\varphi d \in [10,20]$. We also observe that the energy contained in the small scales increases in the vortex outflow region with decreasing radial position.

In summary, should prominent turbulent Taylor rolls exist in the flow, a peak at large scales exists and small-scale structures are present throughout the gap, but most prominently in the vortex ejecting regions. However, when the Taylor rolls have faded, the peak at large scales vanishes and small-scale structures are only detectable close to the cylinder walls. These findings reveal yet another clear evidence of the existence of turbulent non-axisymmetric Taylor rolls and support the idea that the large-scale rolls {consist} of small-scale unmixed plumes. {Furthermore, the azimuthal lengthscale of these plumes is in the order of $k_\varphi d \in [10,20]$.}

\subsection{Spatial correlation coefficients}
As it was shown in the previous section, the pre-multiplied energy co-spectra demonstrate the existence of small-scale structures (coherent plumes) and large-scale azimuthal structures, which are both connected thanks to the presence of turbulent Taylor vortices. For the sake of clarity, the investigation was confined to three specific cases. Within this section, however, we extend the analysis of the large-scale peak based on the azimuthal two-point autocorrelation coefficient. {More specifically, the shift of the large-scale peak close to the cylinder walls as well as its characteristic in the center of the gap are worked out in more detail.} The spatial two-point autocorrelation coefficient of the velocity fluctuations between two points separated by $\Delta \varphi$ in the azimuthal direction is given by

 \begin{align}
 		R_{rr}\left(r,z,\Delta \varphi \right) &=\frac{\langle u_r^* \left(r,\varphi,z,t \right) u_r^* \left(r,\varphi+\Delta \varphi,z,t \right) \rangle_{\varphi,t}}{\langle u_r^{* 2} \left(r,\varphi,z,t \right) \rangle_{\varphi,t}}, \\
 		R_{\varphi \varphi}\left(r,z,\Delta \varphi \right) &=\frac{\langle u_\varphi^* \left(r,\varphi,z,t \right) u_\varphi^* \left(r,\varphi+\Delta \varphi,z,t \right) \rangle_{\varphi,t}}{\langle u_\varphi^{* 2} \left(r,\varphi,z,t \right) \rangle_{\varphi,t}}.
 \end{align}
 
\begin{figure}
  \centerline{\includegraphics[width=\textwidth]{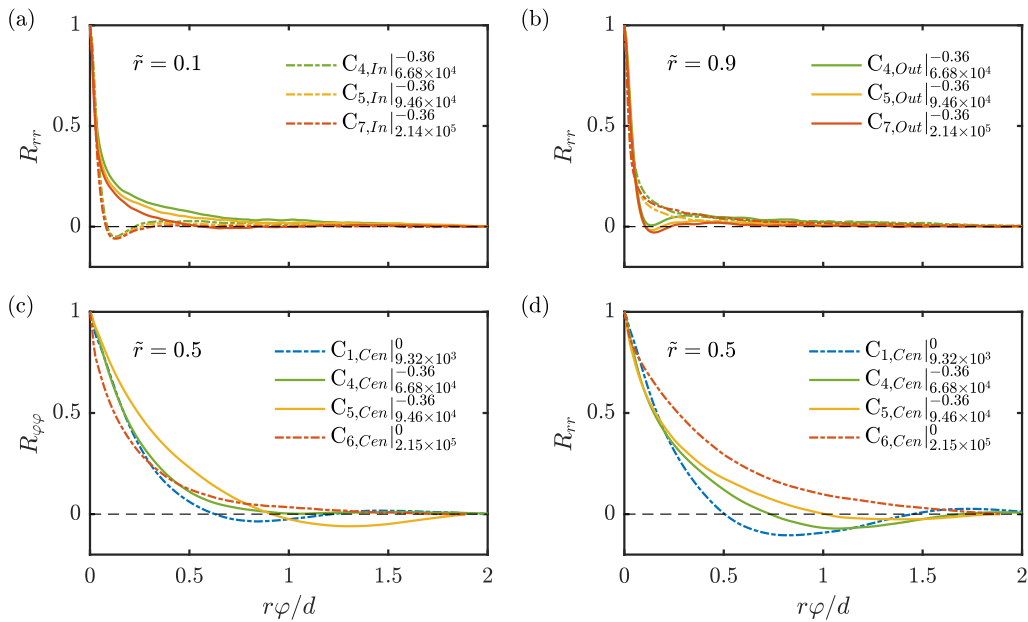}}
  \caption{Upper row: Azimuthal two-point autocorrelation function of the fluctuation radial velocity component at the axial height of vortex in- and outflow at (a) $\tilde{r}=0.1$ and (b) $\tilde{r}=0.9$. Lower row: Azimuthal two-point autocorrelation function of the fluctuation (c) azimuthal and (d) radial velocity component at the axial height of vortex center at $\tilde{r}=0.5$. {Legend abbreviations represent $\text{C}_\# \vert _{Re_S}^{\mu}$}.}
\label{fig:corr_all}
\end{figure}

\noindent In figure \ref{fig:corr_all}a, we show the spatial two-point autocorrelation function of $u_r$ at $\tilde{r}=0.1$ in the ultimate regime for three different flow cases at $\mu_{max}$, and for both the vortex inflow and outflow. We observe that independently of the flow case, the autocorrelation function at the vortex inflow shows a minimum at $r\varphi /d \approx 0.13$, which indicates the existence of azimuthal structures of that size. In contrast, for the outflow, we observe a {monotonic} decrease of the autocorrelation function. Close to the outer cylinder wall (see figure \ref{fig:corr_all}b), the opposite behavior is observed with a pronounced minimum of the autocorrelation function for the vortex outflow at $r\varphi /d \approx 0.15$. This suggests that when the large-scale Taylor rolls transport fluid against the cylinder walls, azimuthal structures seem to be stimulated close to these walls in good agreement with the findings of the spectral analysis shown in $\S$ \ref{sec:Azimuthal energy co-spectrum}. 

Next, in figures \ref{fig:corr_all}c,d, we show both the azimuthal and radial two-point autocorrelation function for the center of the vortex evaluated at $\tilde{r}=0.5$. At this position, the large-scale peak in the co-spectra is most pronounced as it was shown in figures \ref{fig:spectra_CR} and \ref{fig:spectra_UR}. Here, we compare three flow states in the classical and ultimate regime where turbulent Taylor vortices are present and a case where they are absent {($\text{C}_6$)}. These figures reveal that in all vortex dominated cases, a large-scale oscillating behavior is developed, while in {the} absence of rolls, the {autocorrelation} simply decreases towards zero over the whole azimuthal measurement length. We thus conclude, that the large-scale peak in the spectra apparently results from a wavy azimuthal pattern connected to the Taylor rolls.

\section{Complex Proper Orthogonal Decomposition (CPOD)} \label{sec:CPOD}
\subsection{CPOD method}

In order to reveal the flow structure connected to the large-scale oscillation within the flow {found in figures \ref{fig:spectra_CR}, \ref{fig:spectra_UR} and \ref{fig:corr_all}}, we perform a {proper orthogonal decomposition} (POD) {of the velocity fluctuations field}. The POD, also known as {empirical orthogonal function} analysis (EOF), is a technique to extract modes like coherent structures from a flow field that contribute most to the energy of the flow \citep{Taira2017}. It can be further used to identify the most relevant degrees of freedom in a dynamical system. For additional information{,} we refer the reader to the review of \citet{Berkooz1993}. As a classical POD mode cannot capture propagating structures, we perform a complex POD using the Hilbert transform (see \citet{Pfeffer1990}), where a mode is split into two patterns namely its real and its imaginary part with a phase difference of $\pi/2$, based on the algorithm described by \citet{Marple1999}. We further use a combined analysis of both velocity components $u_r$ and $u_\varphi$ for the POD analysis. The applied algorithm is based on three steps \citep{Harlander2011}. {At first we use the Hilbert transform $\mathcal{H}$, to make the velocity fluctuations complex:}
{\begin{align}
u_{r,c}^*&=u_r^*+i\mathcal{H}(u_r^*)\\
u_{\varphi,c}^*&=u_\varphi^*+i\mathcal{H}(u_\varphi^*)
\end{align}}

\noindent {Secondly, the imaginary, transformed velocity fluctuations ($u_{r,c}^*,u_{\varphi,c}^*$) are arranged in a data matrix $\bf D$, where columns are assigned to the spatial grid points ($1:M$) and rows to the time signals ($1:N$), while $M>N$:}

{\begin{align}
{\bf D} &=\begin{bmatrix}
 u_{r,c}^*(x_1,t_1)& \cdots & u_{r,c}^*(x_1,t_N)\\
 \vdots & \ddots & \vdots \\
u_{r,c}^*(x_M,t_1)& \cdots & u_{r,c}^*(x_M,t_N)\\
 u_{\varphi,c}^*(x_1,t_1)& \cdots & u_{\varphi,c}^*(x_1,t_N)\\
 \vdots & \ddots & \vdots \\
u_{\varphi,c}^*(x_M,t_1)& \cdots & u_{\varphi,c}^*(x_M,t_N)
\end{bmatrix}
\end{align}}

\noindent {At the end, a singular value decomposition (SVD) of the data matrix is performed as $\textbf{D}=\bm{\Upphi\Upsigma\Uppsi}^T$. The columns of the matrix ${\bm{\Upphi}}$ contain the left singular vectors of $\textbf{D}$ and therefore the $N$ complex modes $\textbf{CPOD}(r,\varphi)$. The diagonal elements of $\bm{\Upsigma}$ hold the singular values, whose square is a measure of the kinetic energy captured by the individual modes. Note that the singular values are sorted in descending order together with the complex modes, meaning that the first mode represents a larger fraction of kinetic fluctuation energy of the full field than the second one and so forth. Next, the velocity fluctuations at a specific axial location $z$ can be reconstructed by} 

\begin{equation}
\mathbf{u}^*(r,\varphi,t)= \sum_{i=1}^N \text{a}_\text{i}(t) \text{\textbf{CPOD}}_i(r,\varphi).
\end{equation}

\noindent The {time-dependent} coefficients $\text{a}_\text{i}(t)$ result from a projection of the complex modes onto the data matrix, i.e. $\text{a}_\text{i}=\text{\textbf{D}}^T \text{\textbf{CPOD}}_\text{i}$. In the following, we will present the CPOD results for two flow states at the height of the vortex center, where the azimuthal two-point-correlation showed a large-scale oscillating behavior. The first case is in the classical regime at $\mu=0$ and the second one in the ultimate regime at $\mu_{max}$.

\subsection{CPOD in the classical regime}

\begin{figure}
  \centerline{\includegraphics[width=\textwidth]{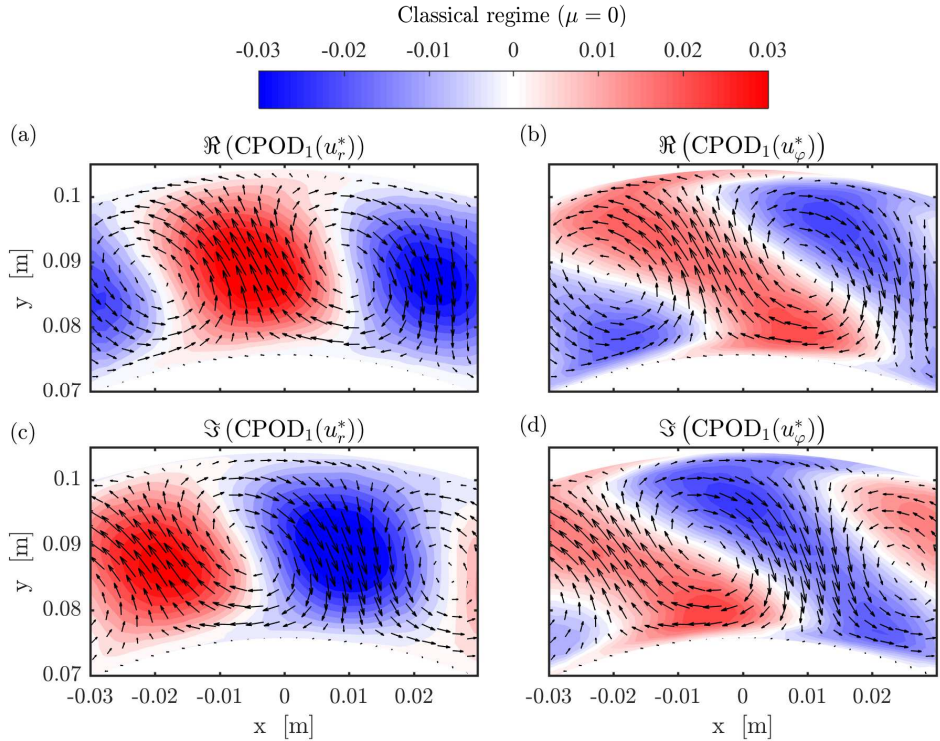}}
  \caption{$\text{\textbf{CPOD}}_\text{1}$ for $\Rey_S=9.32 \times 10^3$ and $\mu=0$ {($\text{C}_1$)} at the axial height of the vortex center for both velocity components. (a) Real and (c) imaginary part of $\text{\textbf{CPOD}}_\text{1}$ for the radial velocity component. (b) Real and (d) imaginary part of $\text{\textbf{CPOD}}_\text{1}$ for the azimuthal velocity component. Black arrows represent the resulting velocity field of $\text{\textbf{CPOD}}_\text{1}$.}
\label{fig:CR_CPOD_Mode1}
\end{figure}

The first CPOD mode for $\Rey_S=9.32 \times 10^3$ and $\mu=0$ {($\text{C}_1$)} is depicted in figure \ref{fig:CR_CPOD_Mode1} for both velocity components. With the first temporal coefficient $\text{a}_1(t)$ and the first mode $\text{\textbf{CPOD}}_\text{1}$, the corresponding velocity fluctuation field is reconstruct  by ${\mathbf{u}_1^*(r,\varphi,t)=\text{a}_1(t) \text{\textbf{CPOD}}_\text{1}(r,\varphi)}$. In figure \ref{fig:CR_CPOD_Mode1}a, the real part of $\text{\textbf{CPOD}}_\text{1}$ representing the radial velocity component, shows nearly circular regions of positive and negative velocity, alternating in the azimuthal coordinate direction. This azimuthal wave pattern becomes azimuthally shifted in the corresponding imaginary part in figure \ref{fig:CR_CPOD_Mode1}c, indicating an azimuthally traveling wave. The real part of the azimuthal velocity of $\text{\textbf{CPOD}}_\text{1}$ in figure \ref{fig:CR_CPOD_Mode1}b, depicts diagonal bands with pointy edges close to the cylinder walls. These regions again show alternating positive and negative velocities. The corresponding imaginary part in figure \ref{fig:CR_CPOD_Mode1}d is also azimuthally shifted, showing the same pattern. As a result, it is revealed that the velocity field consists of counter-rotating vortices in the horizontal plane, whose vorticity axes are co-axial to the rotation axis of the system. The pattern appears similar to that of the well known wavy Taylor vortices (WTV) at lower Reynolds number TC flow. Thus, here we show that in the classical turbulent regime, turbulent Taylor vortices can also feature azimuthal waves.

\begin{figure}
  \centerline{\includegraphics[width=\textwidth]{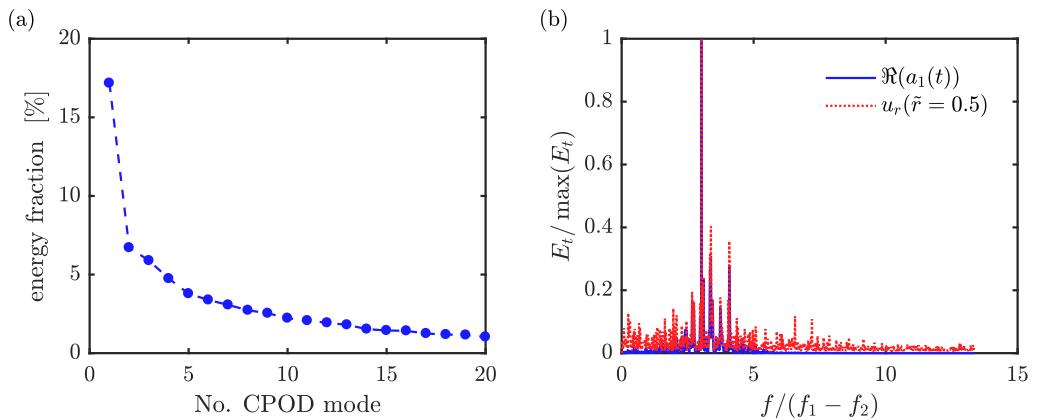}}
  \caption{(a) Turbulent energy fraction captured by the CPOD modes for the case $\Rey_S=9.32 \times 10^3$ and $\mu=0$ {($\text{C}_1$)} at the axial height of the vortex center. Only the first 20 of the total 1500 modes are plotted. (b) Temporal power spectrum of the real part of the temporal coefficient $\text{a}_1(t)$ of $\text{\textbf{CPOD}}_\text{1}$ and of the radial velocity component of the full field at $\tilde{r}=0.5$, averaged over $\varphi$. The frequency $f$ is normalized by the difference of the cylinder frequencies $f_1-f_2$.}
\label{fig:CR_CPOD_PC_f}
\end{figure}

In figure \ref{fig:CR_CPOD_PC_f}a, we show the energy fraction captured by the CPOD modes for $\Rey_S=9.32 \times 10^3$ and $\mu=0$ {($\text{C}_1$)}. The first mode captures approximately 17\% of the total energy fluctuation and we observe a strong drop to the second mode down to 6.7\%. Accordingly, the first CPOD mode is dominant and is, therefore, the only one discussed. The temporal energy {spectrum} of the real part of the first mode's temporal coefficient $\text{a}_1(t)$ is pictured in figure \ref{fig:CR_CPOD_PC_f}b, together with the temporal energy {spectrum} of the full field radial velocity component, calculated at $\tilde{r}=0.5$ and averaged over $\varphi$. The dominant frequency of $\text{\textbf{CPOD}}_\text{1}$ is $f/(f_1-f_2)=3.03$, identical to the one of the radial velocity component. Thus we can assume that the first mode captures the temporal behavior of the full field quite accurately. The frequencies $f_1$ and $f_2$ represent the rotation frequencies of the inner and outer cylinder, respectively. 

\begin{figure}
  \centerline{\includegraphics[width=\textwidth]{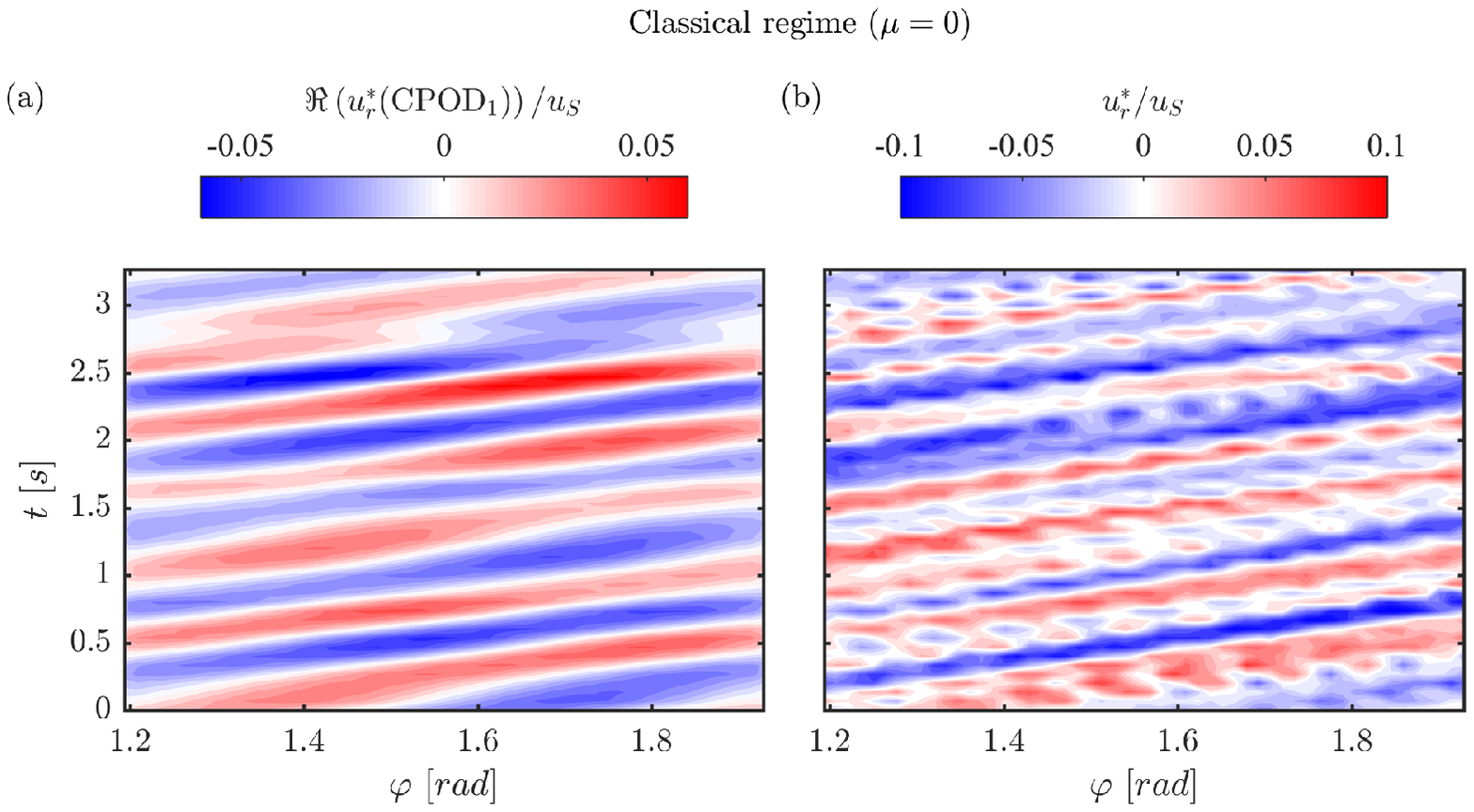}}
  \caption{Space-time diagram of (a) the reconstructed radial velocity component based on the first CPOD mode and (b) the full field radial velocity component for $\Rey_S=9.32 \times 10^3$ and $\mu=0$ {($\text{C}_1$)} at the axial height of the vortex center as function of $\varphi$ and $t$. The radial coordinate is fixed at $\tilde{r}=0.5$.}
\label{fig:CR_CPOD_spacetime}
\end{figure}

To illustrate the {spatiotemporal} character of the first mode, we depict in figure \ref{fig:CR_CPOD_spacetime}a its reconstructed radial velocity component as well as the corresponding full field at $\tilde{r}=0.5$ as a function of the azimuthal coordinate $\varphi$ and time $t$. The space-time plot of $\text{\textbf{CPOD}}_\text{1}$ depicts diagonal bands of alternating positive and negative velocity, representing azimuthally propagating waves into the direction of the mean flow. Furthermore, the same diagonal bands -superimposed by turbulent fluctuations- are observed in the full field. A reappearance of azimuthal waves for pure inner cylinder rotation at a similar radius ratio of $\eta=0.733$ has already been found by \cite{Wang2005} in the Reynolds number regime of $20 \leq \Rey/\Rey_C \leq 38$. In our study in the classical regime for $\mu=0$ at $\eta=0.714$, we find $\Rey_S/\Rey_{S,C} \approx 99$, where the critical shear Reynolds number is located around $\Rey_{S,C}\approx94.5$ \citep{Ostilla14b}. This provides evidence of the reappearance of an azimuthal wave in the classical turbulent regime for a much larger turbulence intensity. Moreover, we can extract the corresponding wave frequency $f_w$, wavelength $\lambda_w$ and phase speed $c_w$ of $\text{\textbf{CPOD}}_\text{1}$. The temporal phase function $\phi_i$ is defined as

\begin{equation} \label{temp_phase}
\phi_i(t)=\arctan\left( \frac{\Im (a_i(t))}{\Re (a_i(t))} \right).
\end{equation}

\noindent Its temporal derivative is equal to the angular frequency of the CPOD mode \citep{Susanto1998}, leading to a wave frequency of $f_{w,1}=\partial_t\phi_1(t)/(2 \pi)=1.80 \ \text{Hz}$. In addition, the spatial phase function $\Phi_i(x)$ is defined as

\begin{equation} \label{spatial_phase}
\Phi_i(r,\varphi)=\arctan\left( \frac{\Im (\text{CPOD}_i(r,\varphi)}{\Re (\text{CPOD}_i(r,\varphi))} \right),
\end{equation}

\noindent {whose azimuthal derivative} is a measure for the local wavenumber $k_{\varphi,w}$. Note that the CPOD analysis yields only one temporal coefficient $\text{a}_i(t)$, but two spatial modes concerning the radial and azimuthal velocity components. Therefore we can extract two wavelengths, which are however nearly identical. The averaged wavelength, evaluated for $\tilde{r}=0.5$, is given by ${\lambda_{w,1}=2 \pi/(\partial_\varphi \Phi_1(\tilde{r}=0.5,\varphi))=0.68\,\text{rad}}$, leading to an azimuthal wavenumber of around $k_{\varphi,w,1}=2\pi/\lambda_{w,1} \approx 9$. In this way, the phase speed becomes ${c_{w,1}=\lambda_{w,1} f_{w,1}=1.22\,\text{rad/s}}=0.35\omega_1$. As the detected wave pattern reminds us of wavy Taylor vortices, we compare our result with the wave speeds for WTV measured by \citet{King1984}, although one should keep in mind the large Reynolds number difference. \citet{King1984} find for the case of pure inner cylinder rotation at $\eta=0.73$, $R/R_c \approx 14$, and an average axial wavelength of $\bar{\lambda}_w/d=2.4$ a wave speed of $c_w \approx 0.2 \omega_1$, while the corresponding pattern features $k_{\varphi,w}=2$. Additionally, they show for a slightly larger radius ratio of $\eta=0.84$, that the wave speed increases when $\Rey/\Rey_c>18$. Thus, we can conclude that the wave speeds superimposed to the Taylor vortices in the wavy Taylor vortex regime and in the turbulent Taylor vortex regime are in the same order. Moreover, the wave speed seems to increase with the forcing $\Rey_S$, as the wave speed in our case is {noticeably} higher than the one reported by \citet{King1984}.

\subsection{CPOD in the ultimate regime}

\begin{figure}
  \centerline{\includegraphics[width=\textwidth]{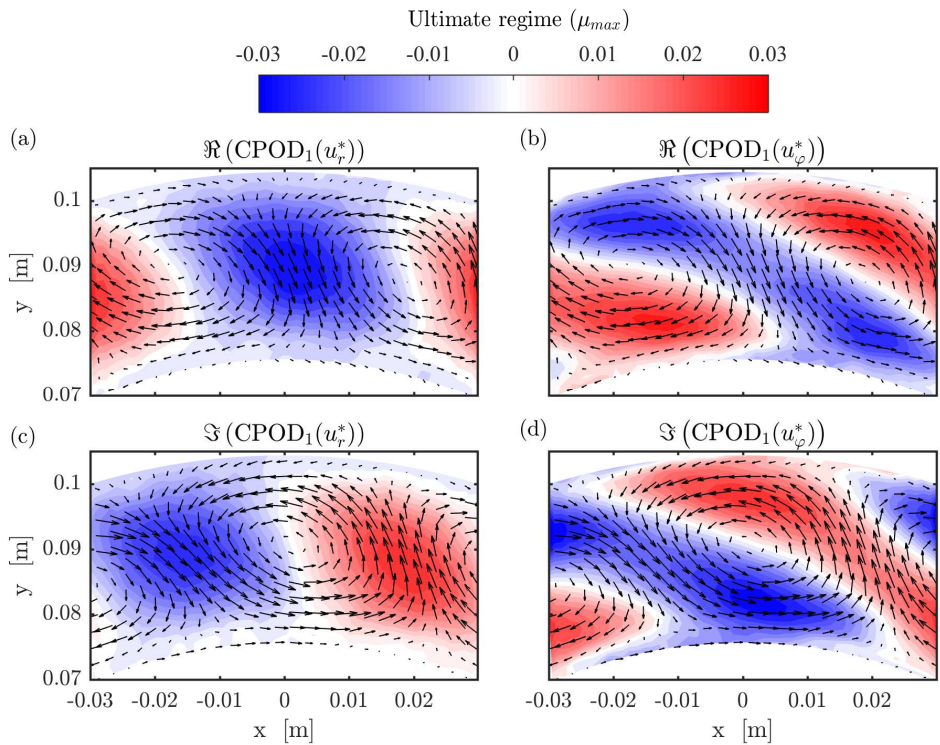}}
  \caption{$\text{\textbf{CPOD}}_\text{1}$ for $\Rey_S=6.68 \times 10^4$ and $\mu=-0.36$ {($\text{C}_4$)} at the axial height of the vortex center for both velocity components. (a) Real and (c) imaginary part of $\text{\textbf{CPOD}}_\text{1}$ for the radial velocity component. (b) Real and (d) imaginary part of $\text{\textbf{CPOD}}_\text{1}$ for the azimuthal velocity component. Black arrows represent the resulting velocity field of $\text{\textbf{CPOD}}_\text{1}$.}
\label{fig:UR_CPOD_1}
\end{figure}

We now turn our attention to the ultimate regime at the torque maximum rotation rate. In figure \ref{fig:UR_CPOD_1}a-d, the first CPOD mode, representing the real and imaginary part of the radial and azimuthal velocity component, for $\Rey_S=6.68 \times 10^4$ and $\mu=-0.36$ {($\text{C}_4$)} at the axial height of the vortex center is depicted. The flow in the ultimate regime features nearly the same pattern as already shown in \ref{fig:CR_CPOD_Mode1} with alternating regions of positive and negative velocity. In case of the radial and azimuthal velocity component, these regions form circular patches and diagonal bands with pointy edges, respectively. In addition, the patterns of the imaginary part are azimuthally shifted relative to the real parts.

{Similar to the previous case in the classical regime, the turbulent energy fraction decreases from the first to the second mode from approximately 12\,\% to 5\,\%. Further, the temporal power spectrum of the temporal coefficient $\Re (a_1(t))$ depicts a prominent peak at $f/(f_1-f_2)=0.92$ (figures are not shown). By use of the temporal and spatial phase functions (see equations \ref{temp_phase} and \ref{spatial_phase}), the azimuthally traveling wave can be described by a wave frequency of $f_{w,1}=3.67\,\textrm{Hz}$, a wavelength of $\lambda_{w,1}=0.74\,\textrm{rad}$ and a phase speed of $c_{w,1}=2.72\,\textrm{rad/s}=0.11\Delta\omega$. Remarkably, the turbulent Taylor vortices feature azimuthal waves also in the ultimate turbulent regime at $\mu_{max}$. To the best of our knowledge, this is the first study reporting such a wave pattern in that highly turbulent regime. It is further worth to mention that a CPOD analysis in the ultimate regime for $\Rey_S=2.15\times 10^5$ and $\mu=0$ {($\text{C}_6$)}, where the large Taylor rolls disappear in the mean field, does not yield any sign of traveling waves.} As a comparison tool, we show in table \ref{table:waves} the results concerning the azimuthally traveling waves found in this study and the findings of \citet{King1984} for $\eta=0.73$.

\begin{table}
  \begin{center}
\def~{\hphantom{0}}
  \begin{tabular}{cccccc}
  &$\eta$& $\Rey_S$ & $\mu$ & $k_{\varphi,w,1}$ & $c_{w,1} \, \textrm{[rad/s]}$ \\
  Classical regime {($\text{C}_1$)}& 0.714 & $9.32 \times 10^3$ & 0 & $\approx 9$ & $0.35 \Delta \omega $ \\
  Ultimate regime {($\text{C}_4$)} &0.714& $6.68 \times 10^4$ & $-0.36$ & $\approx 9$ & $0.11 \Delta \omega$\\
  \citet{King1984} &0.730& $1.05\times 10^3$& $ 0$& $2$ & $0.21 \Delta \omega$ 
  \end{tabular}
  \caption{Overview of detected azimuthally traveling waves in the classical and ultimate regime in comparison to the findings of \citet{King1984}.}
  \label{table:waves}
  \end{center}
\end{table}

To what extent the detection of azimuthal traveling waves is dependent on the turbulence level ($\Rey_S$) and the geometry of the system ($\eta, \Gamma$) is not known.
We note that further experimental as well as numerical investigations would be required to address this issue and reveal whether it is {an} intrinsic property of the flow or not.

\section{Summary \& conclusions}
By means of planar PIV measurements performed in horizontal planes at different axial heights, we showed the dependence of the small-scale statistics and flow organization on the presence of large-scale Taylor rolls in high-Reynolds number TC flow. The ratio of angular velocities $\mu$ is the appropriate parameter to control whether or not the Taylor rolls within the gap are prominent and stable in a statistical sense. While in the classical turbulent regime Taylor rolls are observed for both $\mu=0$ and $\mu<0$, they fade out when the ultimate regime is reached for $\mu=0$, but can be formed again when sufficient counter-rotation ($\mu<0$) is introduced, in particular at $\mu_{max}$ where the flux of angular momentum is maximum. This turbulent state can then be used to compare the flow ---for the same $\Rey_S$--- with and without the presence of these large-scale rolls.

{To uncover the interplay between large-scale rolls and small-scales plumes in a statistical manner, PDFs of both velocity components and the net convective Nusselt number have been evaluated over {cylindrical} surfaces and at specific axial positions. The PDFs of the radial and azimuthal velocity component are close to Gaussian, when evaluated at a fixed height, in accordance with the results of \citet{Huisman2013}. However, when all heights covering one vortex pair are included, the Gaussian shape is only preserved for $\mu=0$ for both velocity components but changes drastically for $\mu_{max}$. There, the PDFs of $u_\varphi$ feature a cusp-like shape with exponential tails, which is a fingerprint of intermittent small-scale plumes and can also be found in RB flow concerning the temperature \citep{Emran2008,Brauckmann2016b}. The PDFs of $\text{Nu}_\omega^{c,net}$ fluctuate around zero and not the averaged value of the Nusselt number and can be described (for $\mu=0$) {by the shape of a distribution of the product of two Gaussian variables}. At $\mu_{max}$ however, the right-hand tail broadens with increasing $\Rey_S$, which can be attributed to an increasing number of rare and strong events of momentum flux at the axial location of the vortex in- and outflow. These {axially-dependent} events originate from the ejection of coherent plumes from the cylinder walls into the gap and lead to a dominant contribution of the vortex in- and outflow to the overall momentum transport. While the contribution of the vortex inflow reaches a nearly fixed value in the ultimate regime of around $\approx 30\%$ at $\mu_{max}$ in the outer gap region, the contribution of the outflow increases monotonically with $\Rey_S$. Here{,} we find a strikingly large value of the angular momentum transport of $\approx 60\%$ in the inner gap region at $\Rey_S=2.14 \times 10^5$.}

{The energy content and azimuthal lengthscale of these small-scale flow structures in the presence of large-scale rolls are calculated based on azimuthal energy co-spectra. The small-scale plumes show an azimuthal extent of $k_\varphi d \in [10,20]$ and contribute most to the overall fluctuation energy in regions where the large-scale rolls transport fluid away from the wall, i.e. the so-called ejection regions \citep{Ostilla14b}. In addition, the energy spectra combined with a correlation analysis revealed that the large-scale Taylor rolls feature an azimuthally oscillating behavior instead of being axisymmetric and stimulate mid-scale azimuthal structures of sizes $r\varphi/d \approx 0.13-0.15$, where fluid transported by these rolls impact on the cylinder walls.}

{In order to finally capture the underlying flow structure of non-axisymmetric Taylor rolls, we performed a complex POD analysis at the height of the vortex center. The turbulent Taylor vortices are superimposed by large-scale azimuthally traveling waves not only in the classical regime at $\mu=0$, but also in the ultimate regime at $\mu_{max}$. These waves propagate into the direction of the mean flow and are similar to the well-known wavy Taylor vortices. While the wave speeds of $0.35 \Delta \omega$ in the classical, and $0.11 \Delta \omega$ in the ultimate regime are in the same order as the one for the wavy Taylor vortices measured by \citet{King1984}, the azimuthal wave number found of $k_{\varphi,w,1}=9$ is much higher than in the laminar regime.}

{Our} findings reveal the intrinsic statistical relation between structures of different scales: large-scale Taylor rolls and small-scale plumes in high-Reynolds number TC flow. {Their interplay} strongly relies on the specific locations along the vortex (inflow, center or outflow) and since they play a prominent role for the flow organization and the momentum transport, our study underlines the importance of an axial exploration when studying the statistics of turbulent TC flows.

We finally note that we believe that our results are much more general than only holding for turbulent TC flow. Clearly, they will generalize to turbulent RB flow with its organization on very large scales \citep{Stevens2018,Schumacher2018}, but also to pipe and channel flow with highly organized structures in spanwise direction\citep{Smits2011,jimenez2018,Marusic2019}. What mechanism, however, sets the lengthscale of the organization of such superstructures in spanwise direction remains unclear.

\section*{Acknowledgments}
The authors would like to thank G.-W. Bruggert, B. Benschop and M. Bos. for their technical assistance. We gratefully acknowledge financial support by the European High-Performance Infrastructures in Turbulence (EuHIT). This work was also funded by an ERC Advanced Grant and by the MCEC program which is part of the Netherlands Organisation for Scientific Research (NWO). C. Sun acknowledges financial support from the Natural Science Foundation of China under Grant No. 11672156. C. Egbers acknowledges financial support by the DFG (EG100/15-2 and EG100/23-1).

\clearpage


\newpage
\bibliographystyle{jfm}

\end{document}